\documentclass[12pt,preprint]{aastex}
\newcommand{\simle}{\mbox{$\stackrel{<}{_{\sim}}$}}
\newcommand{\simge}{\mbox{$\stackrel{>}{_{\sim}}$}}
\newcommand\arcsecc{\hbox{$^{\prime\prime}$}}

\newcommand\lsun{\hbox{\,L$_\odot$}}

\shorttitle{Keck-IOTA Aperture Synthesis}
\shortauthors{Monnier et al.}

\begin{document}

\title{High-resolution imaging of dust shells using Keck aperture
masking and the IOTA Interferometer}


%% author and affiliation information
%% Note that \email has replaced the old \authoremail comman
%% from AASTeX v4.0. You can use \email to mark an email addres
%% anywhere in the paper, not just in the front matter
%% As in the title, you can use \\ to force line breaks

\author{J. D. Monnier\altaffilmark{1}, R. Millan-Gabet\altaffilmark{2},
P. G. Tuthill\altaffilmark{3},
W. A. Traub\altaffilmark{4}, N. P. Carleton\altaffilmark{4},\\
V. Coud{\' e} du Foresto\altaffilmark{5},
W. C. Danchi\altaffilmark{6},
M. G. Lacasse\altaffilmark{4}, S. Morel\altaffilmark{7},
G. Perrin\altaffilmark{5},
I. L. Porro\altaffilmark{8}, F.~P.~Schloerb\altaffilmark{9}, 
and C.~H.~Townes\altaffilmark{10}
}

\altaffiltext{1}{monnier@umich.edu: University of Michigan Astronomy Department, 
941 Dennison Bldg, Ann Arbor, MI 48109-1090, USA.}
\altaffiltext{2}{Michelson Science Center, California Institute of Technology}
\altaffiltext{3}{University of Sydney, Physics Department}
\altaffiltext{4}{Harvard-Smithsonian Center for Astrophysics, 60 Garden St,
Cambridge, MA, 02138, USA}
\altaffiltext{5}{Observatoire de Paris-Meudon}
\altaffiltext{6}{NASA Goddard Space Flight Center}
\altaffiltext{7}{European Southern Observatory}
\altaffiltext{8}{Massachusetts Institute of Technology}
\altaffiltext{9}{University of Massachusetts, Amherst}
\altaffiltext{10}{University of California, Berkeley}

%\email{Contact: monnier@umich.edu}


\begin{abstract}
We present first results of an experiment to combine data from Keck aperture
masking and the Infrared-Optical Telescope Array (IOTA)
to image the
circumstellar environments of evolved stars with
$\sim$20~milliarcsecond resolution.  
The unique combination of excellent Fourier coverage at short
baselines and high-quality long-baseline fringe data allows us to
determine the location and clumpiness of the inner-most hot
dust in the envelopes, and to measure the diameters of the underlying
stars themselves.  We find evidence for large-scale inhomogeneities in
some dust shells and also significant deviations from uniform
brightness for the photospheres of the most evolved M-stars.
Deviations from spherically-symmetric mass loss in the red supergiant
NML~Cyg could be related to recent evidence for dynamically-important
magnetic fields and/or stellar rotation.  We point out that dust shell
asymmetries, like those observed here, can qualitatively explain the
difficulty recent workers have had in simultaneously fitting the
broad-band spectral energy distributions and high-resolution spatial
information, without invoking unusual dust properties or multiple
distinct shells (from hypothetical ``superwinds'').  This paper is the
first to combine optical interferometry data from multiple facilities
for imaging, and we discuss the challenges and potential for the
future of this method, given current calibration and software
limitations.
\end{abstract}

%\keywords{instrumentation: interferometers --- techniques: interferometric --- stars: AGB and post-AGB --- stars: atmospheres --- circumstellar matter}



%\keywords{instrumentation: interferometers -- techniques: interferometric 
%-- stars: AGB and post-AGB -- stars: atmospheres --
%circumstellar matter}









%\tableofcontents

\section{Introduction}
  
Since the advent of infrared detectors, the classic tool for studying
circumstellar dust shells has been fitting the spectral energy
distributions (SEDs) using radiative transfer models.  This has been
true for stars across the Hertzsprung-Russell diagram, for young stars
still accreting material as well as for evolved stars with their
winds.  The conclusions of these studies are only beginning to be
tested rigorously through high-resolution imaging in the visible and
infrared, using 8-m class telescopes and long-baseline
interferometers.

In this paper, we focus mainly on dust shells around evolved stars.
Almost all evolved star SEDs can be fitted well using a simple
physically-realistic model including
a star and a spherically-symmetric, uniform-outflow dust shell
\citep[e.g.,][]{rrh1982, ivezic96a}.  This success led to initial
confidence in our understanding of mass-loss mechanisms: that dust
condenses at T$\sim$1000-1500~K out of a dense stellar atmosphere with
a scale height larger than hydrostatic, maintained by shocks launched
from photospheric pulsations \citep[see recent reviews
by][]{hearn90,lafon91,habing96}.  This theory makes definite
predictions for what should be observed when high-resolution imaging
can resolve these objects, both in terms of location and nature of the
dust formation and the time evolution as clouds are accelerated away
from the star by radiation pressure.

Although early speckle results of \cite{dyck84} found near-IR dust
shell sizes consistent with expectations (given the limited spatial
resolution), recent higher-resolution imaging and interferometry have
consistently found strong deviations from a simple mass loss
scenario.  The Infrared Spatial Interferometer (ISI) found evidence
for diverse dust shell properties in their survey of 13 stars
\citep{danchi1994}.  More dramatically, recent speckle and aperture
masking images of the carbon star IRC+10216 have revealed
inhomogeneities and asymmetries on stellar scales
\citep{haniff1998,weigelt1998,tuthill2000a}; only a few years earlier, a
spherically-symmetric, uniform-outflow model was successfully fit to
the SED \citep{ivezic96}.  Virtually every recent published attempt to
incorporate high-resolution spatial information into SED models has
led to the conclusion that there are strong deviations from the simple
mass-loss prescription of uniform outflow and spherical symmetry.
\citep[e.g.,][]{monnier1997, lopez1997, hale1997, monnier99a,
wittkowski1998, gauger1999,blocker1999,hofmann2001}, at least for the
most evolved and dust-enshrouded sources.

While SED models are adequate for estimating some basic parameters
about dust shells and mass-loss rates (average optical depth and
temperatures), they can not definitively answer some important
questions regarding dust condensation conditions, grain properties,
and the basic mass-loss mechanisms (e.g., source of wind energy and
atmospheric extension).  High-resolution observations, however, can
potentially answer these questions by imaging dust as it forms and
accelerates away from the star.  This morphology and dynamical
information is much better for constraining the wind and mass-loss
theories.  Current interferometer technology is beginning to provide
this: ``movies'' of the expanding dust shell around IRC~+10216 are
already available \citep{tuthill2000a,weigelt2002}.

Until recently, high-resolution images of dust shells could only be
made of the ``biggest'' sources using aperture masking and speckle
interferometry.  In this paper, we extend the capability to
$\sim$20~milliarcsecond scales by combining Keck aperture masking
data, which samples baselines up to 9~m, with IOTA interferometer
data, which samples out to 38~m.  By constraining the long-baseline
visibility, we are able to make higher fidelity images of the inner
dust shells.  This allows us to measure the inner radius of dust
condensation and to search for signs of dust shell asymmetry and
clumpiness, information critical to validating (or falsifying) our
current theories of mass loss.

Lastly, we want to connect our efforts to image evolved stars with
beginning efforts to image disks around Young Stellar Objects (YSO).
The history of YSO SED modeling is beginning to resemble the history
for evolved stars recounted above.  Interferometry results
\citep{rmg1999a,akeson2000,
millangabet2001,tuthill2001a,akeson2002,monnier2002a, lkha2,
colavita2003} have found profound differences from the predictions of
the ``successful'' disk models based on fitting to SEDs alone
\citep{hillenbrand1992, hartmann1993, chiang1997}. 
The new
high-resolution imaging techniques developed here will soon be applied
to imaging preplanetary disks around young stars using new
interferometer facilities, such as the Center for High Angular Resolution
Array (CHARA).

The organization of this article is as follows. We begin by describing
the nature of the observations and the facilities used to acquire the
high resolution data.  Next, we describe the data analysis,
including the results of extensive validation experiments using
new observations of RT~Vir, R~Leo, R~Hya, and W~Hya (an appendix
details our novel calibration method).  We then discuss the
results on each of the ``dust shell'' targets: HD~62623, IRC~+10420,
VY~CMa, NML~Cyg, VX~Sgr, and IK~Tau. These analyses include diameter
fitting, radiative transfer modeling, and image reconstructions. We
also include a general discussion regarding the difficulties in imaging with
new optical interferometers.

Future papers will take up the challenge of creating self-consistent
2-D or 3-D radiative transfer models of the individual sources.
Considering the increased interest in this area recently, this paper
provides an important and timely dataset for other modellers of
evolved stars and dust shells.

\section{Observations}

In this study, we combine data obtained using aperture masking on the
Keck-I telescope \citep{tuthill2000,monnier99} and using the FLUOR
(Fiber Linked Unit for Optical Recombination) beam combiner
\citep{fluor1998} on the IOTA (Infrared-Optical Telescope Array)
interferometer \citep{traub1998}.  The circumstellar environments of
evolved stars are known to change with time, both due to variable
mass-loss on the many-year timescale \citep[e.g.,][]{haniff1998,
  monnier1997} and due to large-amplitude pulsations on shorter
timescales \citep{danchi1994,perrin1999}.  Thus, coordinated (near
simultaneous) observations at both facilities were deemed critical to
avoid possible changes in dust shell morphology between observations.

Here we report on all dust shell targets of this aperture synthesis
effort except for the carbon star V~Hya, the subject of a separate
paper (Millan-Gabet, in preparation), and Table~\ref{targets}
lists the target sources and their basic properties.
Table~\ref{observations}
contains a full journal of our observations relevant to this paper,
where it can be seen that Keck and IOTA measurements were typically
made within a month of each other.  In some cases (detailed later), we
have also included data from other epochs for comparison.  While all
observations were done inside the astronomical K-band (2.0-2.4$\mu$m),
the Keck data used narrow band filters while the IOTA/FLUOR experiment
used a broad band K$^\prime$ filter; this and other factors lead to
systematic errors which are discussed fully in section
\S\ref{systematics}.

\subsection{Aperture Masking}

Aperture masking interferometry was performed by placing aluminum
masks in front of the Keck-I infrared secondary mirror.  This
technique converts the primary mirror into a VLA-style interferometric
array, allowing the Fourier amplitudes and closure phases for a range
of baselines to be recovered with minimal ``redundancy'' noise
\citep[e.g.,][]{baldwin86,jennison58}.  For this work, we used both a
non-redundant ``Golay'' mask and a circular ``Annulus'' mask; this
information, along with observing dates, filter bandpasses, and 
calibrator sources, is included in Table~\ref{observations}.  
Aperture mask specifications, implementation description, and
detailed observing methodology can be found in
\citet{tuthill2000} and
\citet{monnier99}. 

For these observations the Near InfraRed-Camera
\citep{ms94,matthews96} was used in a fast readout mode, adopting an
integration time of 0.137\,s per frame.  Some of the data were corrupted
by highly-variable seeing and ``windshake,'' which blurs the fringes
during the integration time and frustrates precise calibration.  In
situations where suspect calibration is indicated by our data pipeline
diagnostics, previous (and/or subsequent) epochs of data have been
included as a cross-check against possible faulty calibration.  These
individual situations are discussed on a case-by-case base later in
the paper.

\subsection{IOTA-FLUOR}
Long-baseline observations described in this paper were carried out at
the Infrared-Optical Telescope Array (IOTA), a Michelson stellar
interferometer located on Mount Hopkins, Arizona \citep[see][for a description of
the IOTA instrument at the time of these observations]{traub1998}.
Observations were made in the near-IR K$'$ ($\lambda_0 = 2.16 \mu m,
\Delta \lambda = 0.32 \mu m$) bands using three different IOTA
configurations, with physical telescope separations between $B = 21$ m
(North/South orientation) and 38 m (N-NE/S-SW orientation). For
reference, the resolution corresponding to the longest baseline, as
measured by the full-width at half maximum (FWHM) of the response to a
point source, is $\lambda/2B = 6$ mas at K$'$.

The IOTA observations reported here (2000 February, April, June) were
all carried out using the FLUOR beam combiner \citep[Fiber Linked Unit
for Optical Recombination,][]{fluor1998} which uses single-mode fibers
as spatial filters to achieve better precision in the measurement of
fringe visibilities than achievable with bulk-optics combiners.  A
single-mode fiber essentially converts phase errors, caused by
atmospheric turbulence and aberrated optics, into amplitude
fluctuations which can be monitored and corrected
\citep{shaklan1987,shaklan1989}.  In FLUOR, the light from each
telescope is fed into a fluoride glass fiber and split into two parts.
One part is directly sent to the detector as a monitor of the flux
coupling efficiency (``photometric'' signal), while the other is used
for interference in a fiber coupler.  By using the photometric
signals, the fringe visibility can be precisely normalized for each
measurement, thus calibrating effects of varying atmospheric
turbulence.  The fringes are modulated on the detector by a scanning
piezo mirror placed in one leg of the interferometer, a
fringe-detection scheme referred to as temporal modulation.

A typical single observation consisted of 200 scans obtained in $\sim$4~min,
followed by calibration measurements of the background and
single-telescope fluxes (important for characterizing the fiber
coupler chromatic response).  Target observations are interleaved with
an identical sequence obtained on an unresolved or partially-resolved
star, which serves to calibrate the interferometer's instrumental
response and the effect of atmospheric seeing on the visibility
amplitudes.  The target and calibrator sources are typically separated
on the sky by 5-10 degrees and are observed a few minutes apart; these
conditions ensure that the calibrator observations provide a good
estimate of the instrument's transfer function.  The high
brightness of our targets necessitated using similarly bright
calibrators, which were partially resolved on the longest baselines.
Uncertainty in the sizes of these calibrators dominate the calibration
error in most cases, and we have compiled a list of the adopted
angular diameters and sizes in Table~\ref{calibrators}.

\section{Data Reduction}
\label{reduction}
After briefly describing the basic data reduction procedures, we 
will present the results of validation experiments.

\subsection{Keck Aperture Masking}
\label{masking}
The analysis procedures for extracting the visibility amplitudes and
closure phases are well-documented in \citet{tuthill2000} and
\citet{monnier99}.  When performing image reconstructions, the Maximum
Entropy Method (MEM) \citep{sb84,mem83} has been used to create
diffraction-limited images from the interferometric data, as
implemented in the VLBMEM package by \citet{sivia87}.  Other
engineering and performance details may be found in
\citet{tuthill2000} and \citet{monnier99}, while
other recent scientific applications of the data pipeline can be found in
\citet{monnier2002} and \citet{lkha2}.

\subsection{IOTA-FLUOR}
Reduction of the FLUOR data was carried out using custom software
developed using the {\em Interactive Data Language} (IDL), similar in
its main principles to that described by \citet{foresto1997}.
Significant efforts were made to validate the new data pipeline, and
these are detailed in \S\ref{validation}.

Here we briefly summarize the main steps in the data reduction
procedure.  We have included a more detailed description in
Appendix~\ref{appendixa}, including an explanation of our novel
normalization scheme (\S\ref{normscheme}).  Our data pipeline includes
data inspection, determination of a ``kappa'' matrix to characterize
transfer function of fiber optics beam combiner, removal of
photometric fluctuations, fringe amplitude normalization, power
spectra measurement, calibration of instrumental response by observing
calibrator stars, and standard data quality checks. Most targets were
observed multiple times and the visibility measurements showed good
internal consistency from night-to-night.

\subsection{Systematic Errors}
\label{systematics}
The most significant systematic errors in this experiment come from
the aperture masking data at Keck (i.e., not from IOTA-FLUOR).  In
order to have reasonably-low read noise, limitations of the NIRC
camera electronics restrict the integration time of each ``speckle''
frame to $\ge$0.137~s, many times longer than the typical
atmospheric coherence time at 2.2$\micron$ ($\sim$40~ms).
Even worse is
``windshake'' that occurs when observing low elevation sources into
the wind, a common problem with large-aperture telescopes
which results in a blurring of the fringes. 
Most damagingly, this can induce asymmetric mis-calibrations
which must be carefully guarded against.  Miscalibrations are usually
identifiable in the raw data, thus allowing corrupted data to be
flagged.  In cases where we suspect problems (due to
obvious windshake before or after the target), we have included
previous/subsequent epochs of data as a cross-check, or have
limited our analysis to the azimuthal-averages of the visibility data.

Fortunately, fringe-blurring problems have virtually no effect on the
measurements of the closure phases, which remain well-calibrated and
are crucial to the imaging process when the image is not
centro-symmetric. In addition, the excellent Fourier coverage of the
Keck masking allows
hundreds of visibility points to be measured simultaneously, allowing
averaging to recover high precision even when individual baselines
show large fluctuations due to fringe-blurring (as long as
due to statistical fluctuations of normal seeing -- a systematic error
occurs when wind-shake is present).

One common calibration difficulty encountered with the Keck aperture
masking can be empirically corrected.
When the
coherence length $r_0$ or coherence time $t_0$ varies between
observing the source and its calibrator, the overall ratio changes
between the fringe power and the total flux on the detector.
Fortunately for aperture masking data, this change is nearly constant
as a function of baseline, for baselines longer than the coherence
length ($\sim$0.5m at K band).  In practice, this means the observed
visibility function will approach a non-unity visibility at short
baselines (e.g., $V_0=1.05$).  As long as there is no significant flux
coming from large scales ($\sim$0.5\arcsecc, a reasonable assumption
at these wavelengths, but not strictly true due to scattering by
dust), we can renormalize the visibility and recover reasonable data quality
($\simle$10\% visibility errors on the longest baselines).

We have applied an empirical correction (simple scaling) for each epoch of
aperture masking data before combining with IOTA data. 
An overall scaling of the visibility does not usually
affect the image reconstruction process, but can here because we are
combining the Keck data with IOTA-FLUOR results.  We have chosen to
apply this ``correction'' to all the Keck data rather than be selective; usually
this correction is only a few percent, but is occasionally larger.
Data will be presented both with and without this correction.
The calibration factor was arrived at by fitting a Gaussian to the visibility data
for baselines shorter than 1.5~m and using the derived y-intercept extrapolated
to zero baseline.

In contrast to the relatively poor visibility calibration of the Keck
aperture masking data, the IOTA-FLUOR experiment can
produce visibility measurements with $<$1\% precision, under some
circumstances \citep{perrin1999,perrin2003}.  Achieving this precision
requires control of many possible systematic errors, including
corrections for chromaticity, detector non-linearities, and
bandwidth-smearing.  However, this level of precision is not necessary
in this experiment for many reasons.  First, the Keck aperture masking
data typically suffers from greater (5-10\%) calibration errors due to
the effects discussed above, which fundamentally limits our analysis.
Second, our sources have relatively low visibility fringes, meaning
our IOTA measurements are photon-noise limited (or limited by
knowledge of the calibrator stellar sizes) and not limited by
systematic errors in most cases.  Third, high-resolution structures in
the dust shells are expected at the $\simge$1\% level, but can not be
modelled/imaged without orders of magnitude more data; this acts as a
kind of ``noise'' on the measurement which can not be expected to be
fit by simple models.

As an aside, we expect it to be quite difficult to achieve 1\%
absolute precision for broadband fringe measurements when the source
and calibrator have quite different spectra (as for dust-enshrouded
targets); narrow-band filters and/or low-resolution spectroscopy
should always be used for precision visibility measurements.  Hence,
while we do not claim $<$1\% precision here, we do validate in the
next section that our precision is $\simle$3\% based on internal
consistency checks and comparison with stars with previously measured
diameters.

\subsection{Validation}
\label{validation}

Tables~\ref{observations} \& \ref{calibrators} contain the observing
and calibrator information for sources observed as part of our
validation experiments, including RT~Vir, R~Leo, W~Hya, and R~Hya.
Originally, the last three were observed to measure the limb-darkening
on these sources, but this has been deemed impossible due to the
limitations in Keck calibration.  However, because the limb-darkening
effects are relatively subtle, these sources still serve to illustrate
the calibration consistency and precision.

\subsubsection{Internal consistency of IOTA-FLUOR visibilities}
\label{internal_consistency}
\begin{figure}[thb]
\begin{center}
%\epsscale{.5}
\includegraphics[angle=90,height=3in]{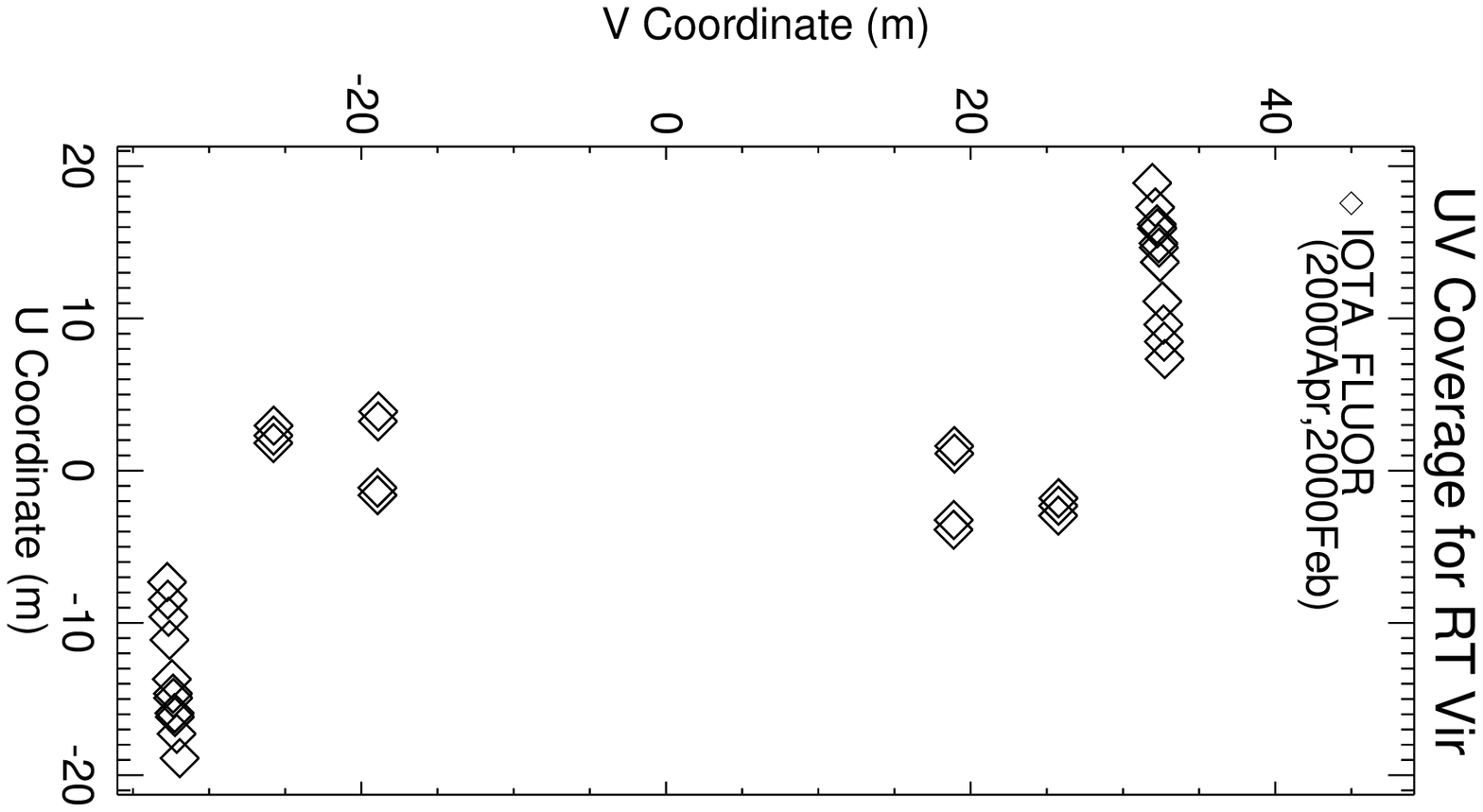}
\includegraphics[angle=90,height=3in]{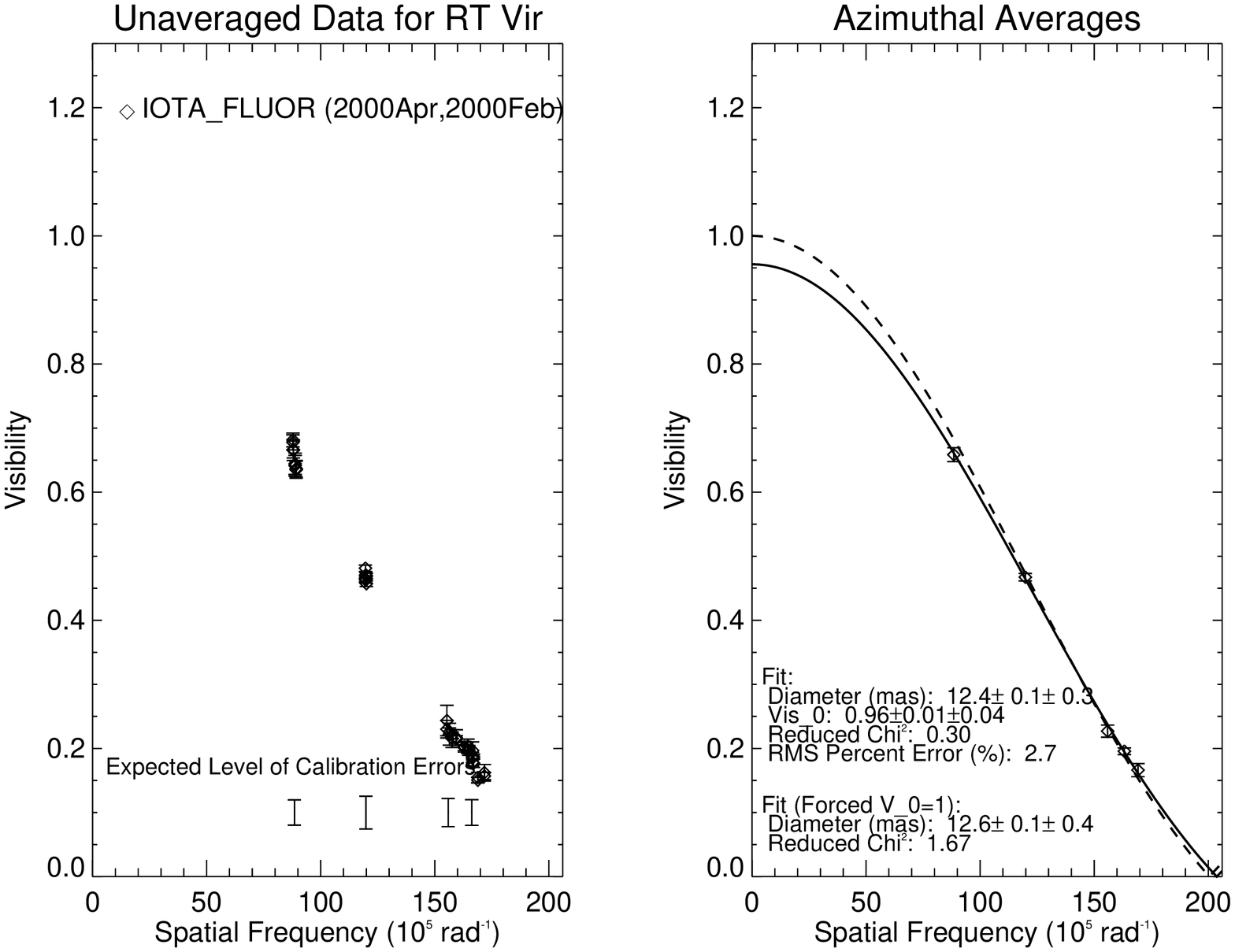}
%\plottwo{f1a.eps}{f2a.eps}
%      \plotfiddle{Figure/fig:astrometry.eps,rot=90}^M
%\plotfiddle{Figures/fig:astrometry.eps}{12cm}{1}{1}{0}{0}{-90}^M
\figcaption{a) UV coverage of the RT Vir observations with IOTA-FLUOR. b)
Unaveraged data for RT Vir.  
Error bars at the bottom of this panel show the expected level of
calibration errors, due to uncertainties in calibrator diameters (which
generally dominate at long baselines) and unmodelled chromatic effects
(which become important for high visibilities); see 
\S\ref{internal_consistency} for the detailed estimation procedure.
c) Averaged data with uniform disk fits.
Here, we present fits with both V(0) fixed to 1.0 and also left free
to vary.  See text \S\ref{validation} for discussion.
\label{figrtvir}}
\end{center}
\end{figure}

First, we validate the new IOTA-FLUOR data pipeline by showing
visibility data of a single source at 3 different baselines. This range
of baselines allows us to check the internal consistency of the data
since the observations involved many configuration changes probing
different resolutions.  Figure~\ref{figrtvir} show the (u,v) coverage
and visibility data for RT~Vir.  Although in general we will be
showing averaged visibility data, here we present each individual
visibility measurement (and error) in Figure~\ref{figrtvir}b.  In this
panel, we also show the expected calibration errors based on the
uncertainty in the calibrator diameters.  In rare cases when the
calibrator uncertainties are not significant, we have assumed a floor
of 3\% systematic error that might arise from unmodelled chromatic
effects (based on software simulations of maximum miscalibrations
possible from strong chromatic differences between source and
calibrator using a model of the FLUOR coupler).

We have fit uniform disk models to this data and have separately
calculated the statistical and systematic errors.  \citet{perrin2003}
presented a sophisticated analytical method for handling this
situation in interferometry data analysis.  An alternate method,
employing bootstrap \citep{bootstrap} and Monte Carlo sampling, is
used here.  For determining the statistical error, random subsets of
averaged data (from each facility) are generated and a best-fit
diameter is calculated for each case. Variance in the fit parameters
directly yield the statistical errors and this method does not require
assumptions concerning the noise distribution.

For the systematic errors, we have used a Monte Carlo method to vary
the sizes of the calibrators used, given the uncertainties from
Table~\ref{calibrators}.  Usually systematic error slightly dominates
over random error in this experiment, although neither affect the
estimated sizes dramatically because the targets are generally heavily
resolved.  Note that all averaging occurs using the original $V^2$ and
not the $V$ in order to avoid bias for noisy datasets; however, we
prefer to present our results using $V$ (which is fully equivalent,
since the errors are small after averaging).

The fit to the RT~Vir data acts as a ``Truth Test'' for our data
analysis pipeline.  The visibility data span a range of 0.2 to 0.7,
allowing a robust test of calibration.  Figure~\ref{figrtvir}c shows
the result of fitting a Uniform Disk (UD) to the dataset, both
allowing the visibility at the origin ($V_0$) to float or be fixed to
unity.  For these two cases, we found the diameter to be
$12.4\pm0.1\pm0.3$~and $12.6\pm0.1\pm0.4$ (the two error estimates are
for statistical and systematic errors respectively, following standard
convention).  Most importantly, the reduced $\chi^2\simle 1$ indicates
a high level of internal consistency to the data calibration.  Our
measurement is similar to the second of the two diameters reported by
\citet{perrin1996}: 13.06$\pm$0.15 mas or 12.36$\pm$0.27 mas,
depending on data selection (see Perrin thesis for more detailed
discussion on this particular source).

When $V_0$ is not fixed to unity, a slightly better fit is found with
$V_0=0.96$.  This slight deviation from a perfect uniform disk could
be due to many plausible mechanisms other than miscalibration,
including changes in photospheric size between 2000Feb and 2000Apr,
non-UD photospheric profile for this late-type star (M8III,
semi-regular pulsator), or a small amount of scattered light from the
known circumstellar dust shell \citep{hron1997}.  Regardless, we have
shown an internal consistency $<$3\% for our data pipeline. While
the true internal calibration might be better than this, the data
quality starts to be become limited by systematic errors.

\subsubsection{Comparing Keck aperture masking and IOTA-FLUOR data}
While most of our targets are complicated dust shell sources, a few
``simple'' sources can act to test the {\em relative} calibration
between the Keck masking and IOTA-FLUOR data.  This comparison
is important since the the interferometry
methods employed are clearly very different: Keck masking used image
plane combination with narrow bandpass filters while IOTA used a fiber combiner
over a broad wavelength band.

\begin{figure}[thb]
\begin{center}
%\epsscale{.5}
\includegraphics[angle=90,height=3in]{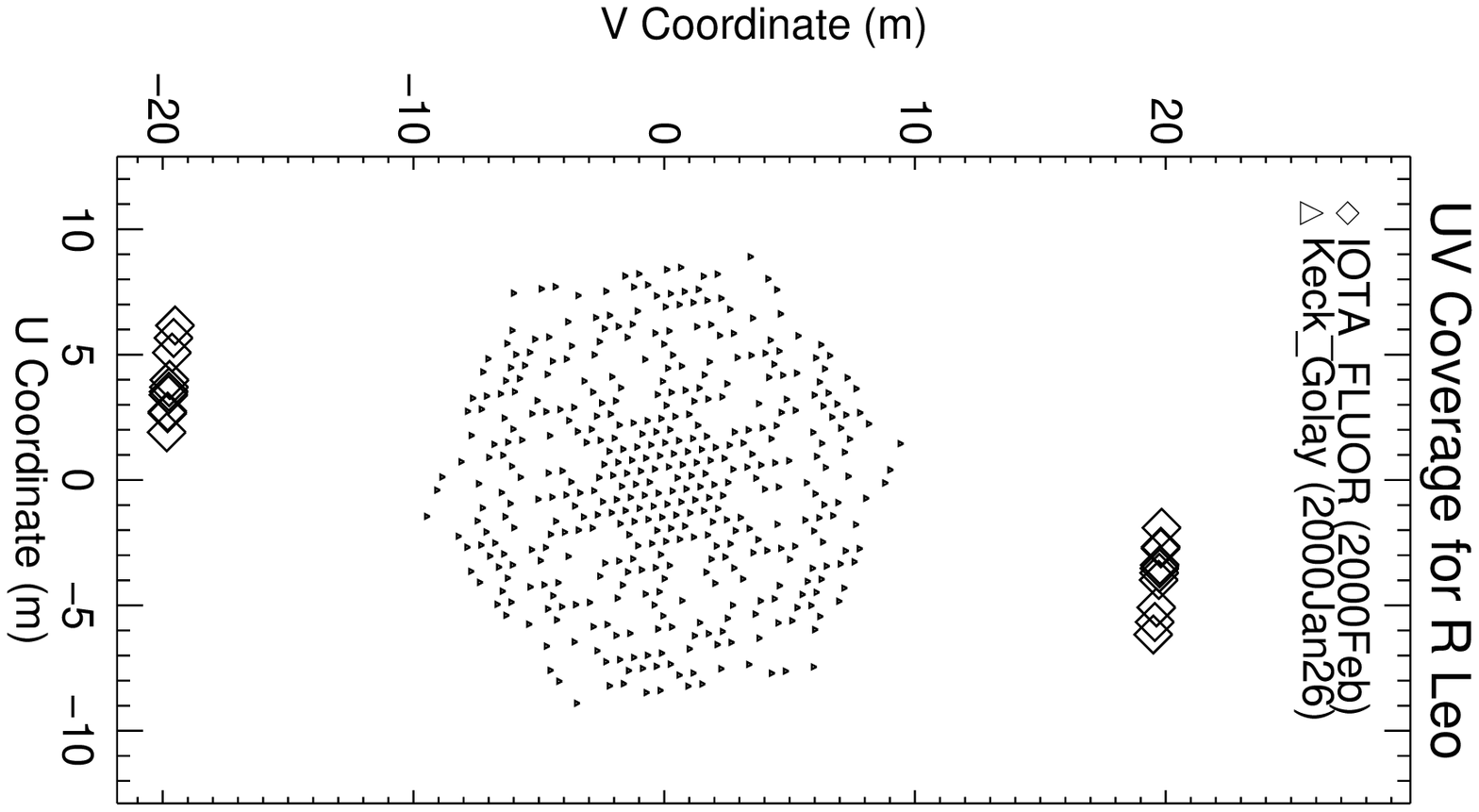}
\includegraphics[angle=90,height=3in]{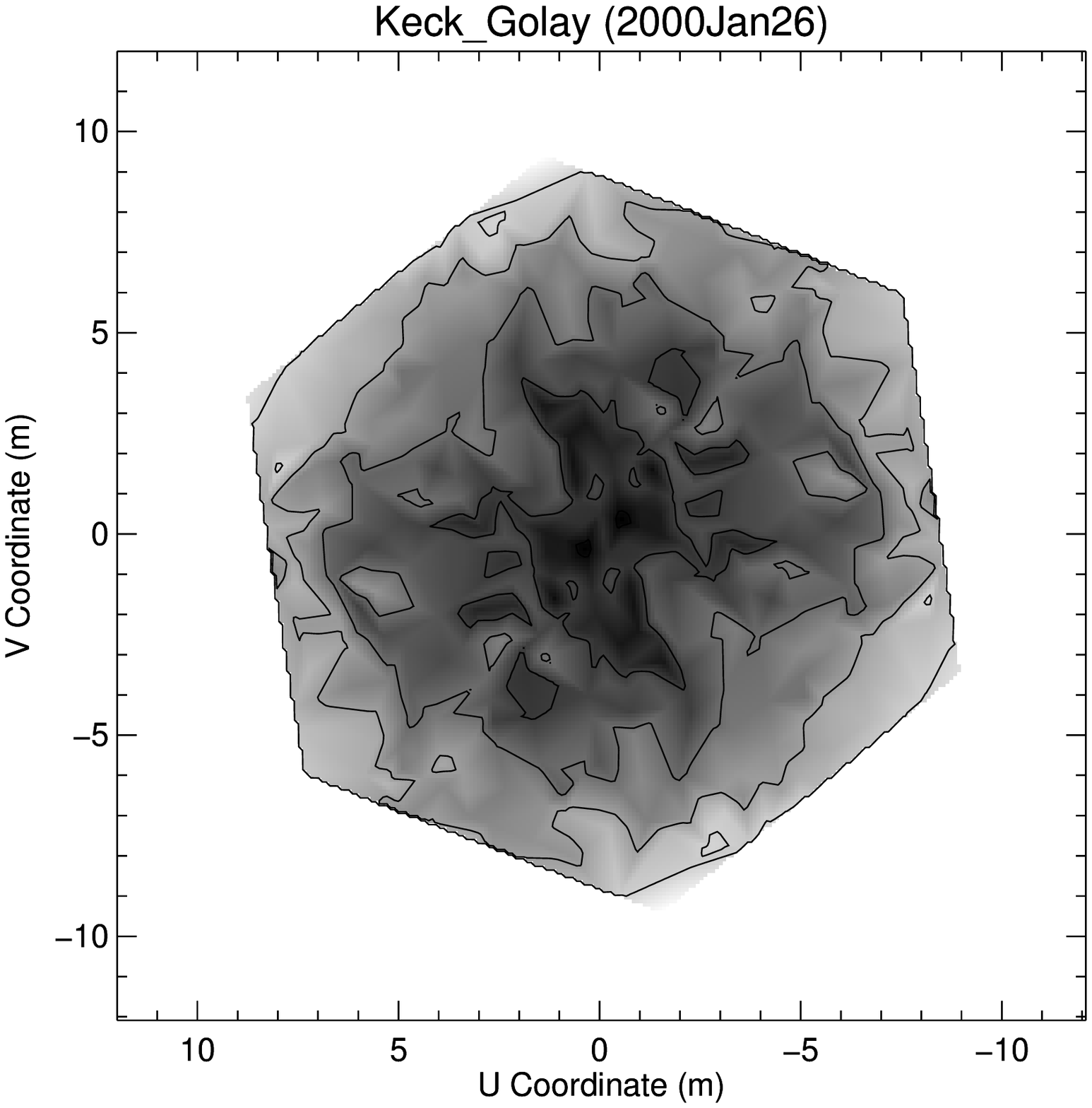}
%      \plotfiddle{Figure/fig:astrometry.eps,rot=90}^M
%\plotfiddle{Figures/fig:astrometry.eps}{12cm}{1}{1}{0}{0}{-90}^M
\figcaption{a) UV coverage of the R Leo Observations from Keck aperture
masking and IOTA-FLUOR.  b) Two-dimensional visibility data observed
using Keck aperture masking. Each contour is 0.1 Visibility.
\label{figrleo1}}
\end{center}
\end{figure}

\begin{figure}[thb]
\begin{center}
%\epsscale{.5}
\includegraphics[angle=90,height=3in]{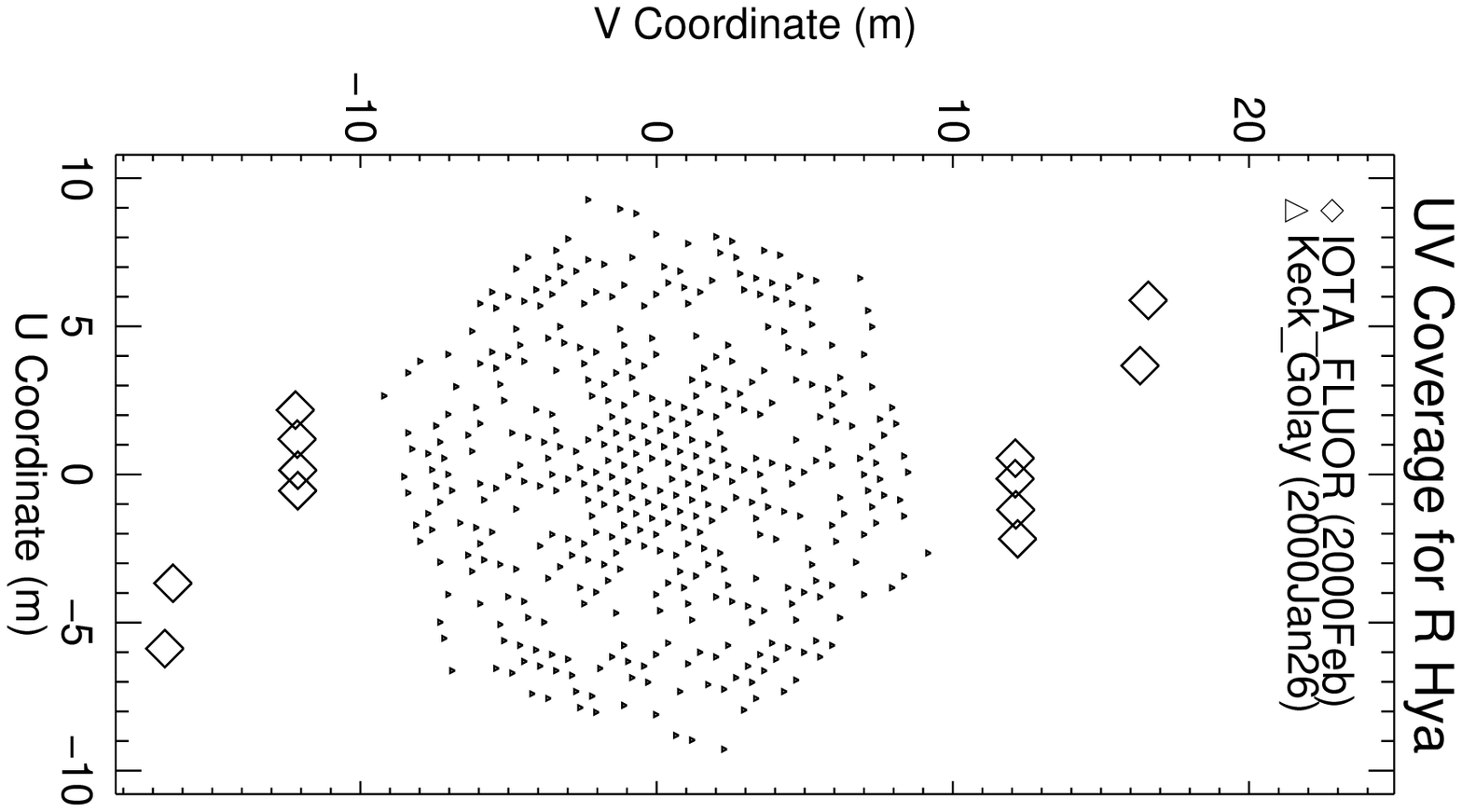}
\includegraphics[angle=90,height=3in]{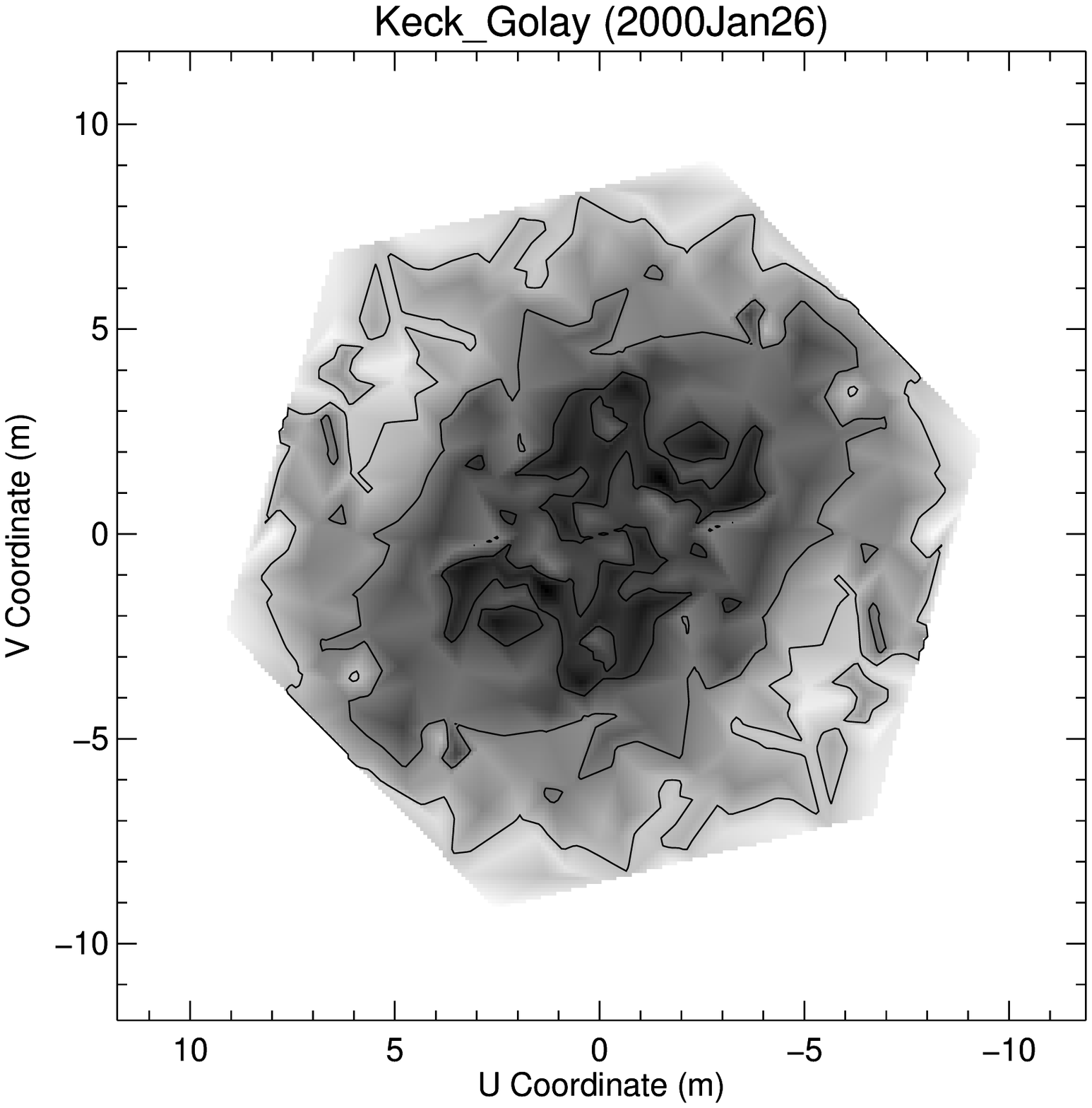}
%      \plotfiddle{Figure/fig:astrometry.eps,rot=90}^M
%\plotfiddle{Figures/fig:astrometry.eps}{12cm}{1}{1}{0}{0}{-90}^M
\figcaption{Same as Figure~\ref{figrleo1}, except for R~Hya.
\label{figrhya1}}
\end{center}
\end{figure}

\begin{figure}[thb]
\begin{center}
%\epsscale{.5}
\includegraphics[angle=90,height=3in]{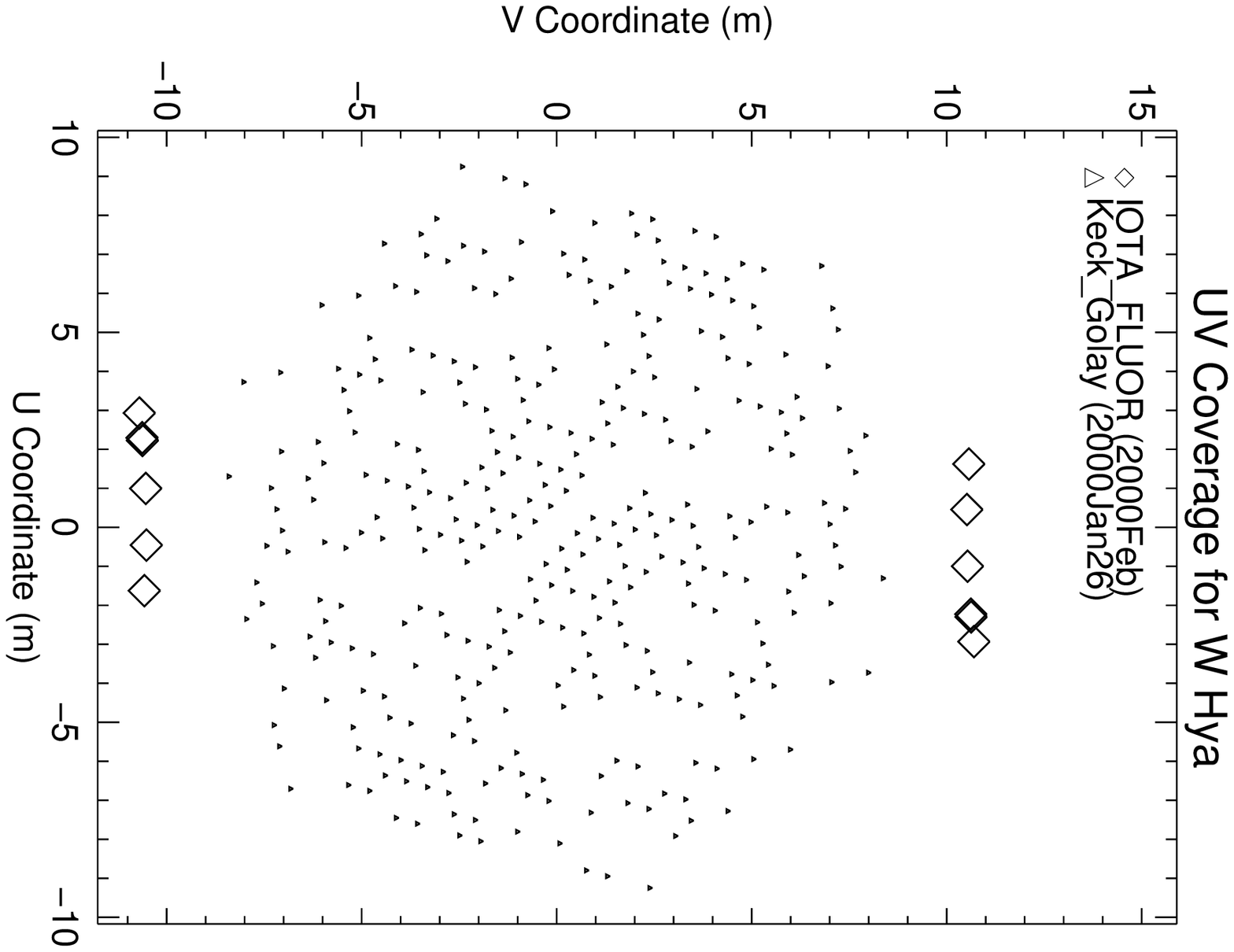}
\includegraphics[angle=90,height=3in]{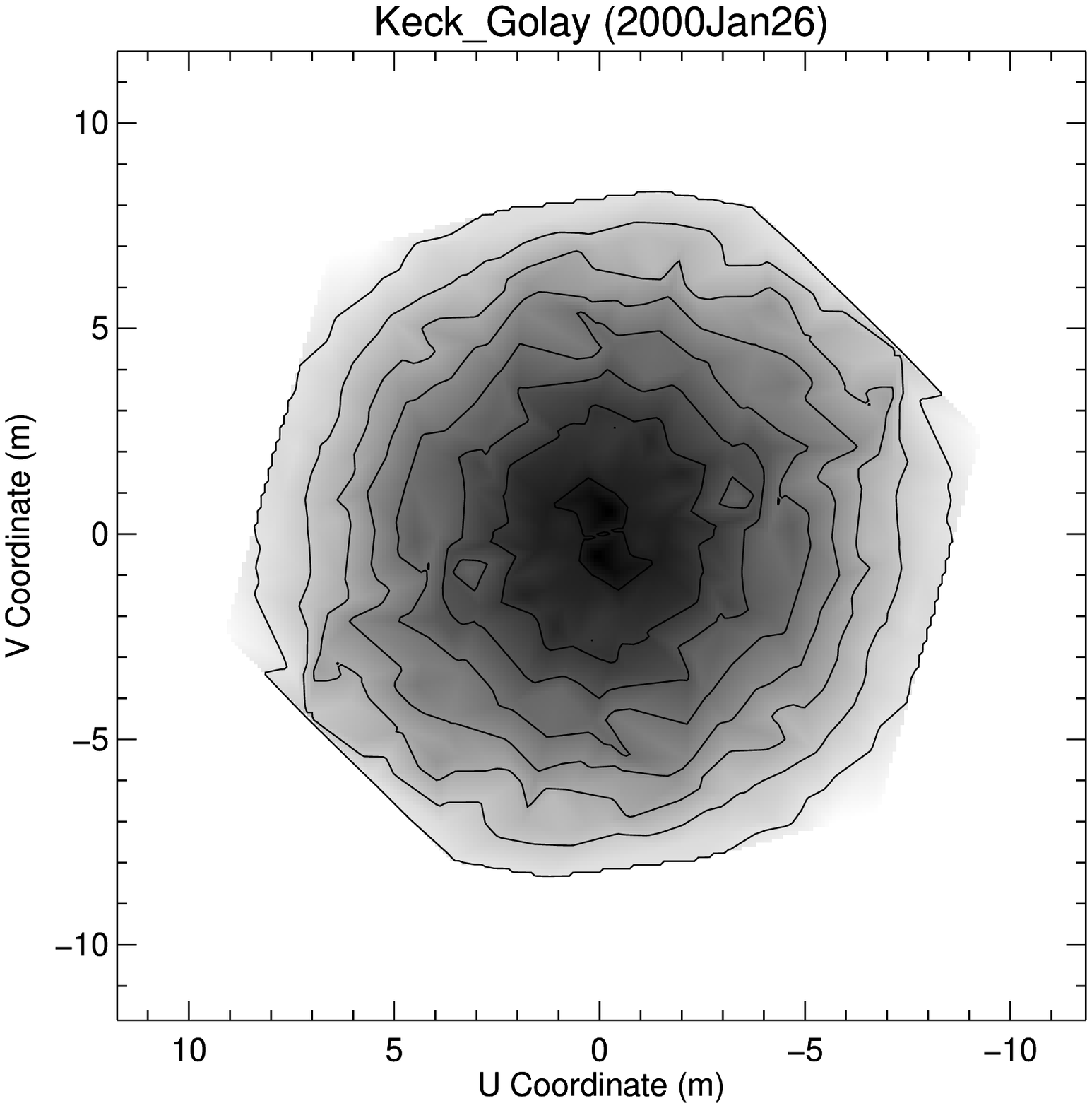}
%      \plotfiddle{Figure/fig:astrometry.eps,rot=90}^M
%\plotfiddle{Figures/fig:astrometry.eps}{12cm}{1}{1}{0}{0}{-90}^M
\figcaption{Same as Figure~\ref{figrleo1}, except for W~Hya.
\label{figwhya1}}
\end{center}
\end{figure}

Figures~\ref{figrleo1}-\ref{figwhya1} contain the (u,v) coverage and
two-dimensional visibility from Keck aperture masking for R~Leo,
R~Hya, and W~Hya.  The low declination of the latter two sources
causes the 21~m physical baseline at IOTA to be projected to
$\sim$11~m, thus providing a near-overlap with the 9~m longest
baselines employed at Keck. This overlap regime allows another good
check of the relative calibration procedures.  The Keck
masking data reveals these sources to be fairly circular, as
expected.  A separate calibrator study has shown that we expect
10-20\% asymmetries from windshake and other systematic errors for
sources of this size (Nick Murphy, private communication, 2003).
Hence, any small residual asymmetries seen are likely to be
miscalibrations and are not modelled here.

\begin{figure}[thb]
\begin{center}
%\epsscale{.5}
\includegraphics[angle=90,height=3in]{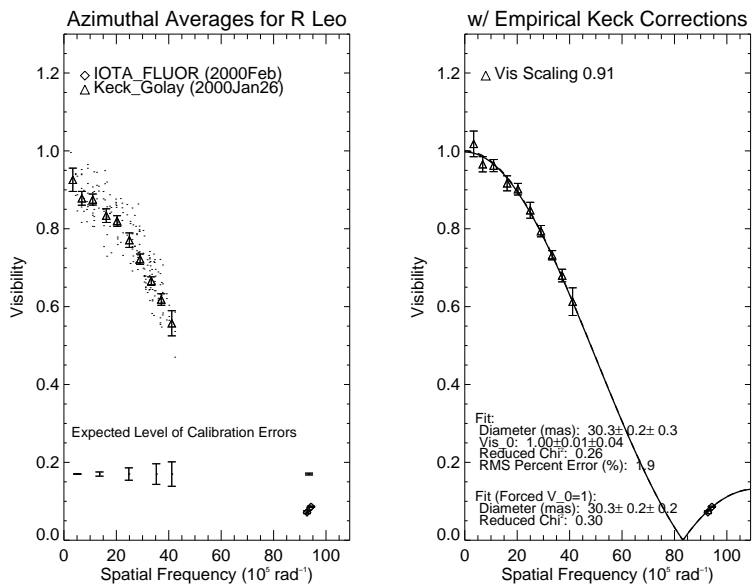}
\figcaption{R Leo Data. a) The left panel shows each individual Keck and
IOTA visibility datum as a point, while the azimuthal averages are
plotted with error bars.  At the bottom, an estimate of the
calibration/systematic errors are shown for each baseline range.  b)
The right panel shows only averaged data and the empirical Keck
corrections have been applied (see \S\ref{systematics}).  
Two curves are shown: solid line is a uniform-disk diameter fit 
where $V_0$ is a free parameter, dashed line is a fit with fixed $V_0$=1.0.
The fitted parameters and reduced $\chi^2$ are included in the legend. 
Two errors are listed for each fitted parameter, corresponding to
the statistical and systematic uncertainties respectively.
(The solid and dashed lines are indistinguishable for these particular
data fits.)
\label{figrleo2}}
\end{center}
\end{figure}

\begin{figure}[thb]
\begin{center}
%\epsscale{.5}
\includegraphics[angle=90,height=3in]{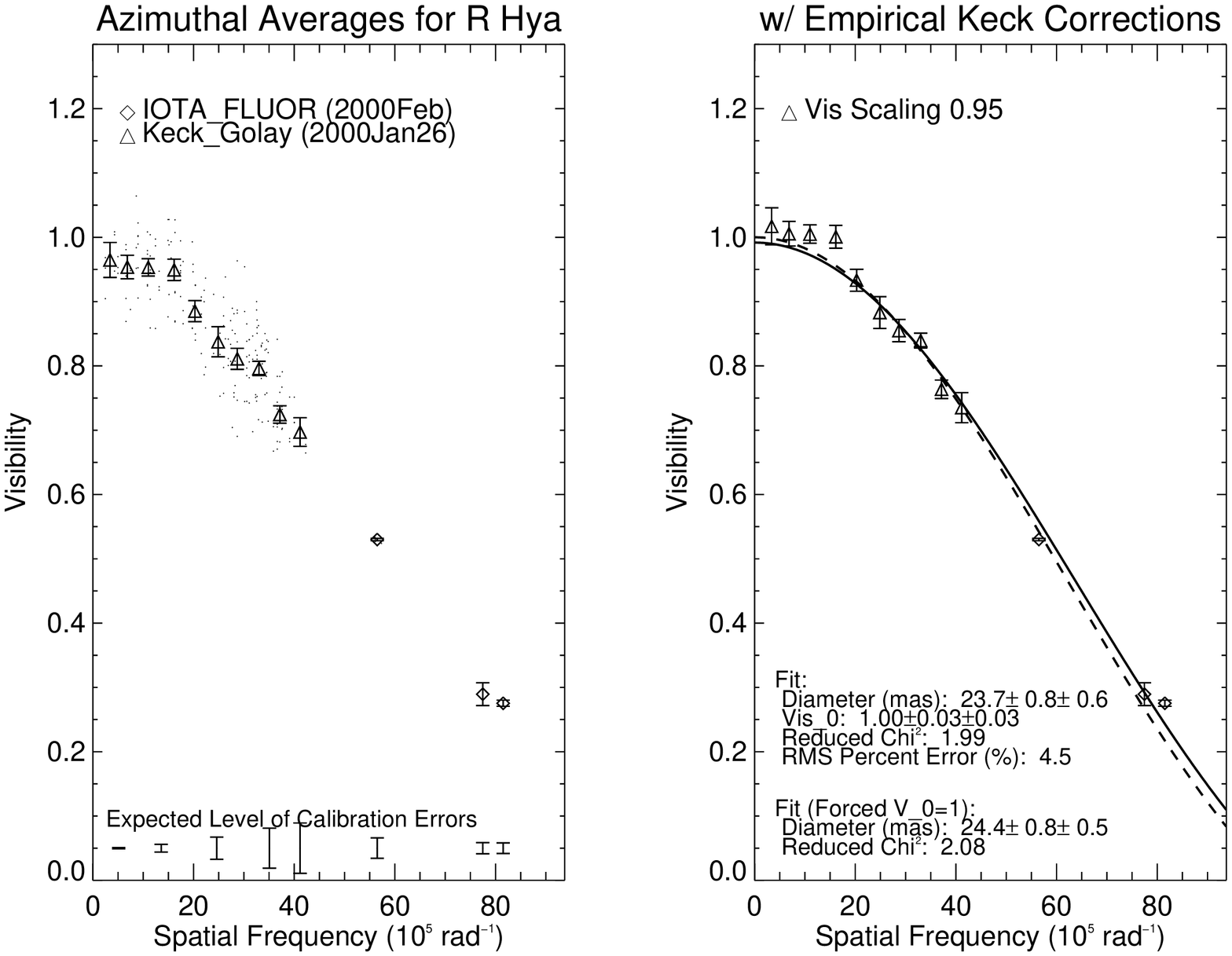}
\figcaption{Same as Figure~\ref{figrleo2}, except for R~Hya.
\label{figrhya2}}
\end{center}
\end{figure}

\begin{figure}[thb]
\begin{center}
%\epsscale{.5}
\includegraphics[angle=90,height=3in]{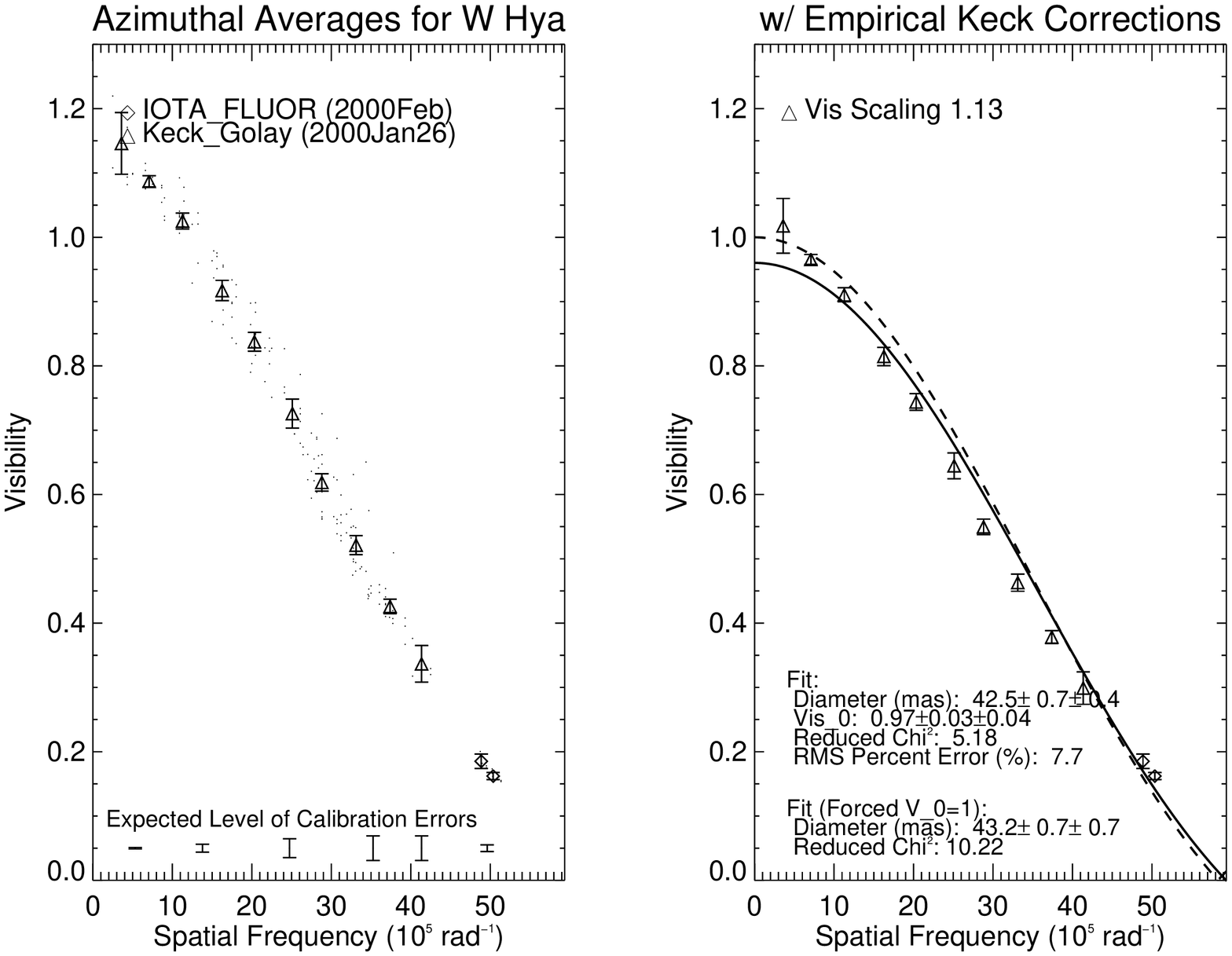}
\figcaption{Same as Figure~\ref{figrleo2}, except for W~Hya.
\label{figwhya2}}
\end{center}
\end{figure}

Figures~\ref{figrleo2}-\ref{figwhya2} show the azimuthally-averaged
data, both before and after applying the empirical Keck corrections described above
in \S\ref{systematics}.  Also, the results of the uniform
disk fits are shown, following the same procedure described above for RT~Vir.
The final results of UD diameter fits and relevant comments can be found
in Table~\ref{diameters}.

R~Leo (Figure~\ref{figrleo2}) shows a surprisingly good agreement at
long and short baselines, completely consistent with a uniform disk
with diameter 30.2$\pm$0.2$\pm$0.3 mas.
This is contrary to recent findings of \citet{perrin1999} who found
strong evidence for deviations from uniform brightness, 
possibly due to (time-variable) molecular opacity effects
\citep[e.g.,][]{mennesson2002, jacob2002}.

The shortest IOTA baselines and the longest Keck baselines are
similar for R~Hya and W~Hya (Figure~\ref{figrhya2} \&
\ref{figwhya2}).  Extrapolations of the Keck visibility show good
agreement, at the $\sim$5-10\% level, consistent with expected
calibration errors of Keck data itself (additional data of HD~62623,
VY~CMa, and IRC~+10420, presented in \S\ref{results} also contain
baseline near-overlaps and confirm this result).  We conclude that any
systematic errors resulting from the use of different filters at Keck
and IOTA are less than other known sources of error. 

R~Hya and W~Hya each show systematic deviations from a uniform disk
profile shown in Figures~\ref{figrhya2} \& \ref{figwhya2}, 
evident from the large reduced $\chi^2$.  While
miscalibration could explain some of the changes, presence of dust
emission and/or molecular layers in the upper atmosphere could also
explain the discrepancies.  Our use of different filter bandpasses for
Keck and IOTA impacts our ability to study this effect, since
molecular absorption dominate near the edges of the K~band
\citep{thompson2002a} and are not probed by the Keck narrowband
filters.  Our observations highlight the need for more systematic
study of Mira photospheres using narrow-band filters or spectroscopy.

Lastly, we consider a few miscellaneous effects which could affect
the absolute data accuracy.  For IOTA-FLUOR, calibration of the
``effective'' wavelength and corrections for bandwidth-smearing can be
important for sources observed at and beyond the first null of the
visibility pattern.  In order to estimate the size of this first
effect, we considered a simple model of the K$^\prime$ filter and a
reasonable range of effective temperatures, finding that the effective
wavelength can only shift by $\sim$1\%.
Bandwidth-smearing destroys any true nulls in a visibility curve,
because only a single wavelength experiences a null for a given
projected baseline (thus non-nulled wavelengths dominate signal when
using a broadband filter).  For IOTA-FLUOR characteristics, the
visibility minimum is $V\sim$2\% at the location of the ``Nulls'' and
the peaks are diminished by $\Delta V \sim 0.003$.  This effect has
been modelled for R~Leo given the specific parameters of these new
observations, and it was found not to significantly change the
diameter estimation above.

\subsection{Data quality summary}
\label{dataquality}
Given that these results represent the first use of IOTA-FLUOR data by
our group and the first combination of it with Keck data, we have
detailed the statistical and systematic errors encountered in a number
of validation experiments.  We have shown that the IOTA-FLUOR data
pipeline is self-consistent with $<$3\% precision in
Visibility for the test observations of RT~Vir.  We find good
agreement between visibilities measured at Keck and IOTA when using
similar baselines on a common source, although detailed testing of
this is currently limited by poor absolute calibration of Keck data at
the longest baselines ($\sim$10\%).  These demonstrations are critical
to give confidence in the fidelity of the results presented in the
next section (and future papers which utilize these data), where 
challenges to data interpretation include
large gaps in the baseline coverage and high uncertainties in
the source models.

\section{Results and discussion}
\label{results}
In this section, we present the visibility data  
for six targets with resolved dust shells:
HD~62623 (A3Iab), IRC~+10420 (F8I), VY~CMa (M3/4I), NML Cyg (M6I), VX
Sgr (M4-9.5I), and IK~Tau (M10III).  The sources are at different
states of stellar evolution and have a range of masses/luminosities.
In this section, we discuss the data, modeling and imaging results
for each source individually.  Detailed modeling will be pursued in future papers using
more sophisticated techniques.

In some cases, simple radiative transfer models were used to interpret
the visibility data and to compare with previous results.  We only
consider spherically-symmetric models consisting of a central star
(with radius $R_\ast$ and effective temperature $T_\ast$) surrounded
by a dust shell with a power-law density profile (usually $\rho\propto
r^{-2}$, uniform outflow).  The dust is assumed to begin at $R_{\rm
inner}$ and extend to $R_{\rm outer}$.  The dust shell optical depth
was parameterized in terms of $\tau_{2.2\mu m}$.  Additional details
regarding the dust optical properties are given on a case-by-case
basis below.

\begin{figure}[thb]
\begin{center}
%\epsscale{.5}
\includegraphics[angle=90,height=3in]{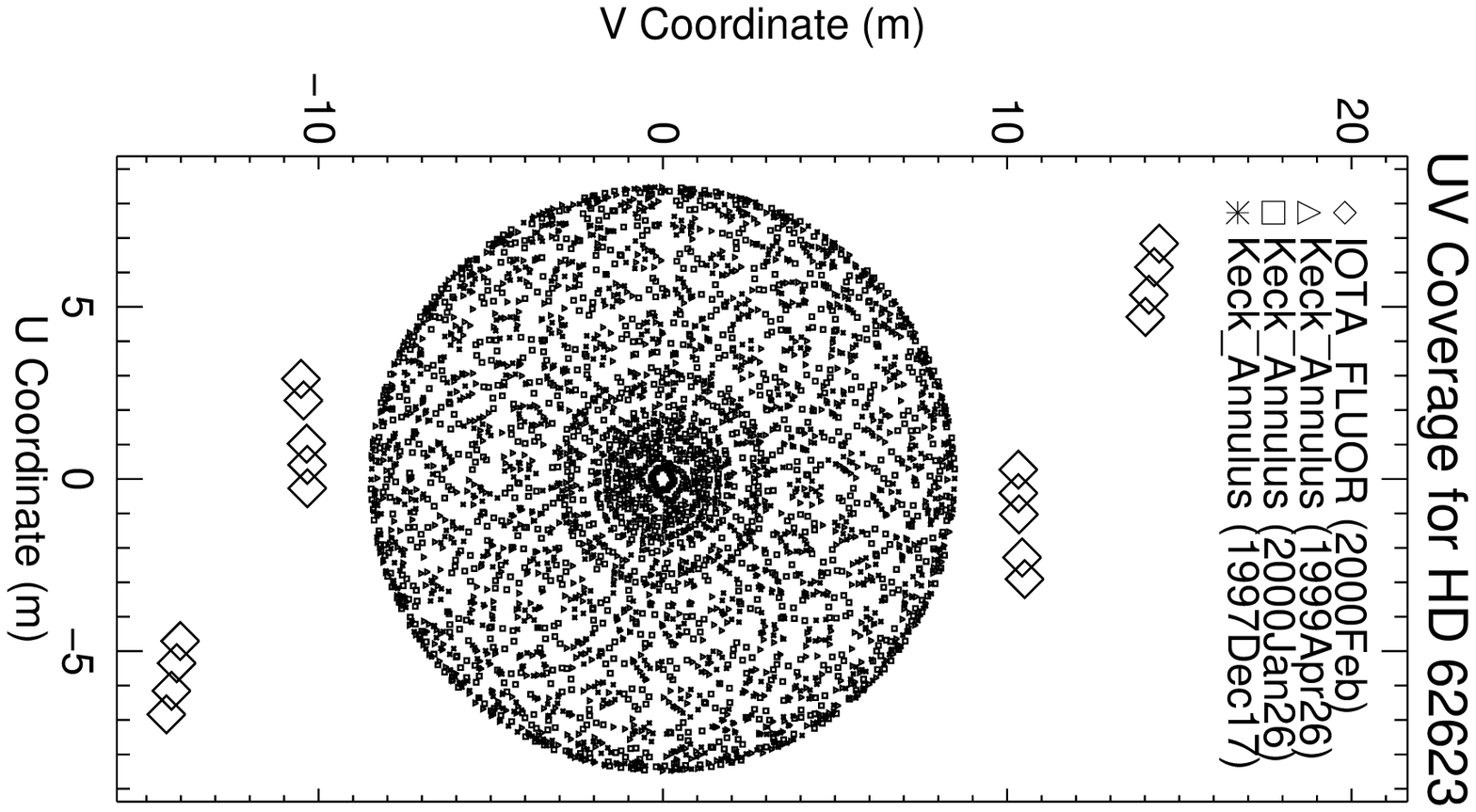}
\includegraphics[angle=90,height=3in]{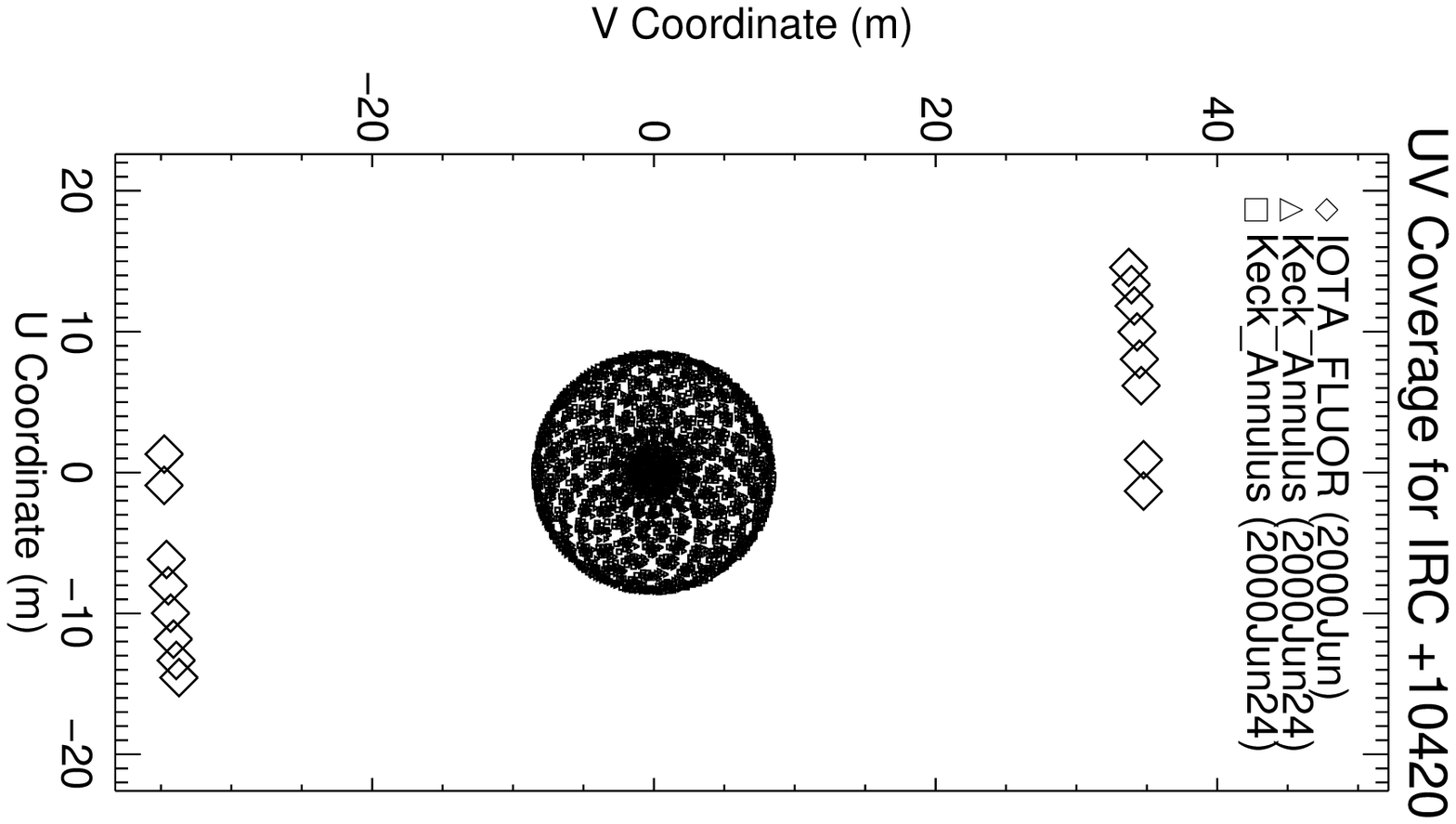}
\includegraphics[angle=90,height=3in]{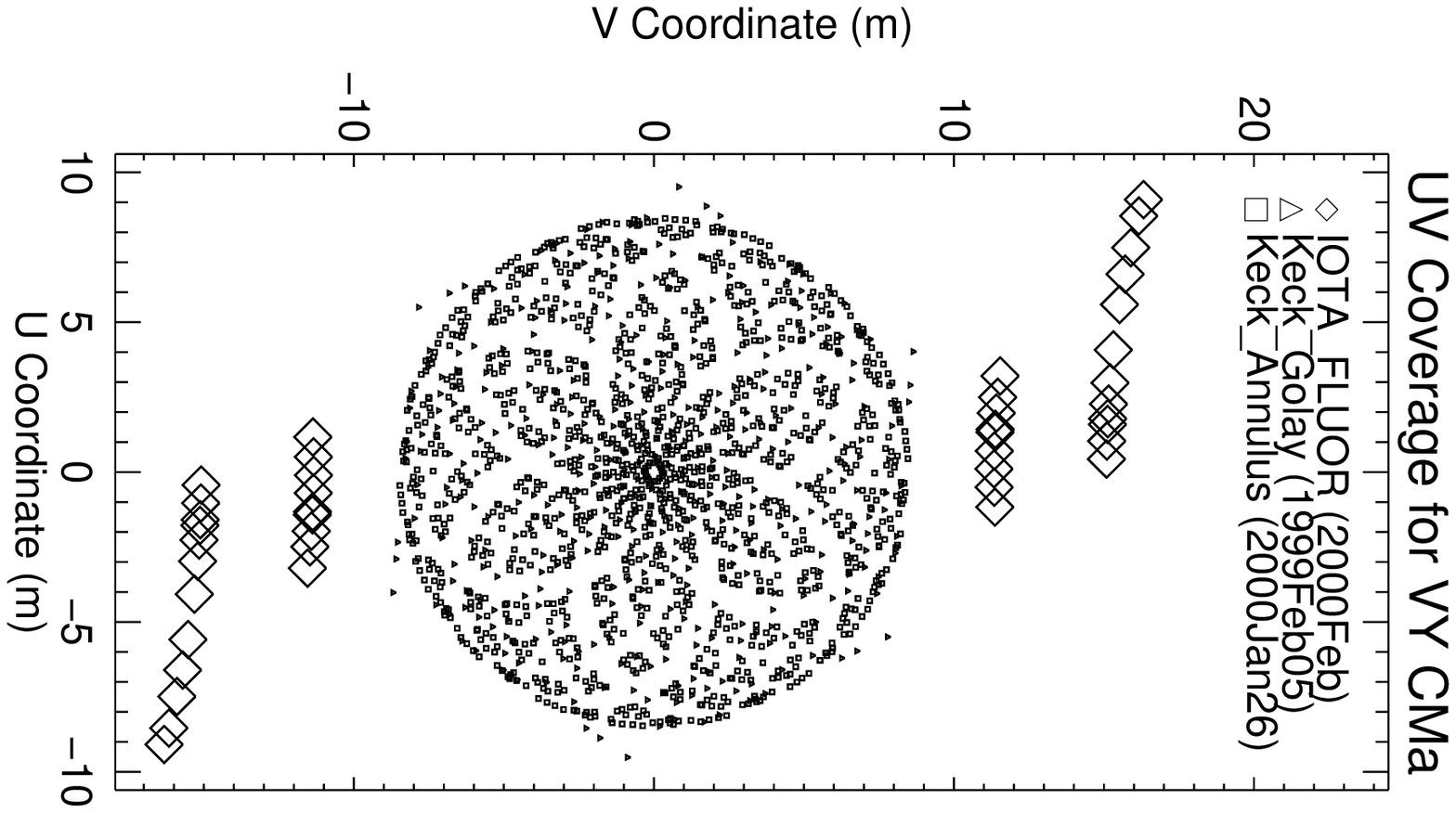}
\figcaption{UV coverage of the new interferometric observations for: a) HD~62623, b) IRC~+10420,
c) VY~CMa
\label{uvcov1}}
\end{center}
\end{figure}

\begin{figure}[thb]
\begin{center}
%\epsscale{.5}
\includegraphics[angle=90,height=3in]{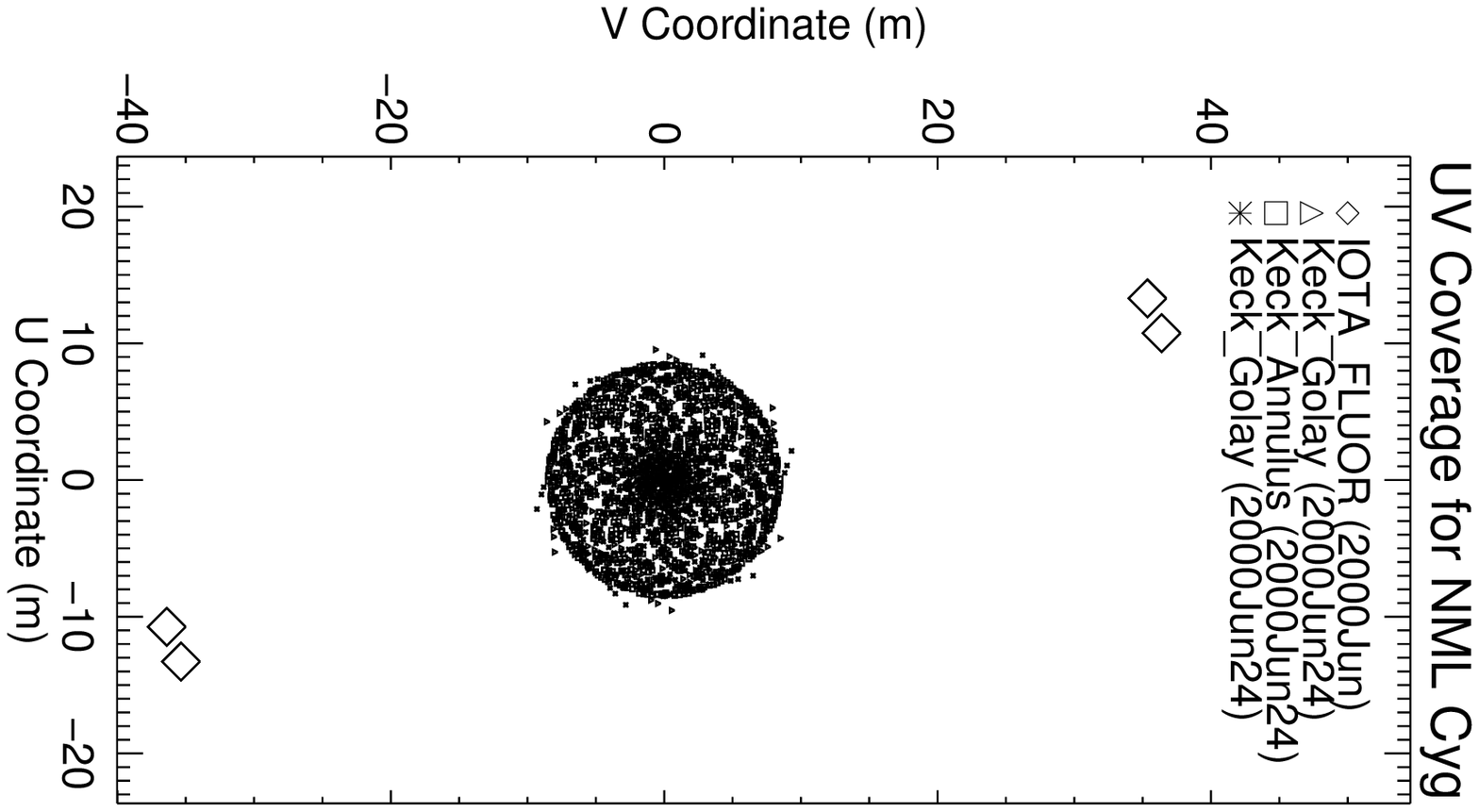}
\includegraphics[angle=90,height=3in]{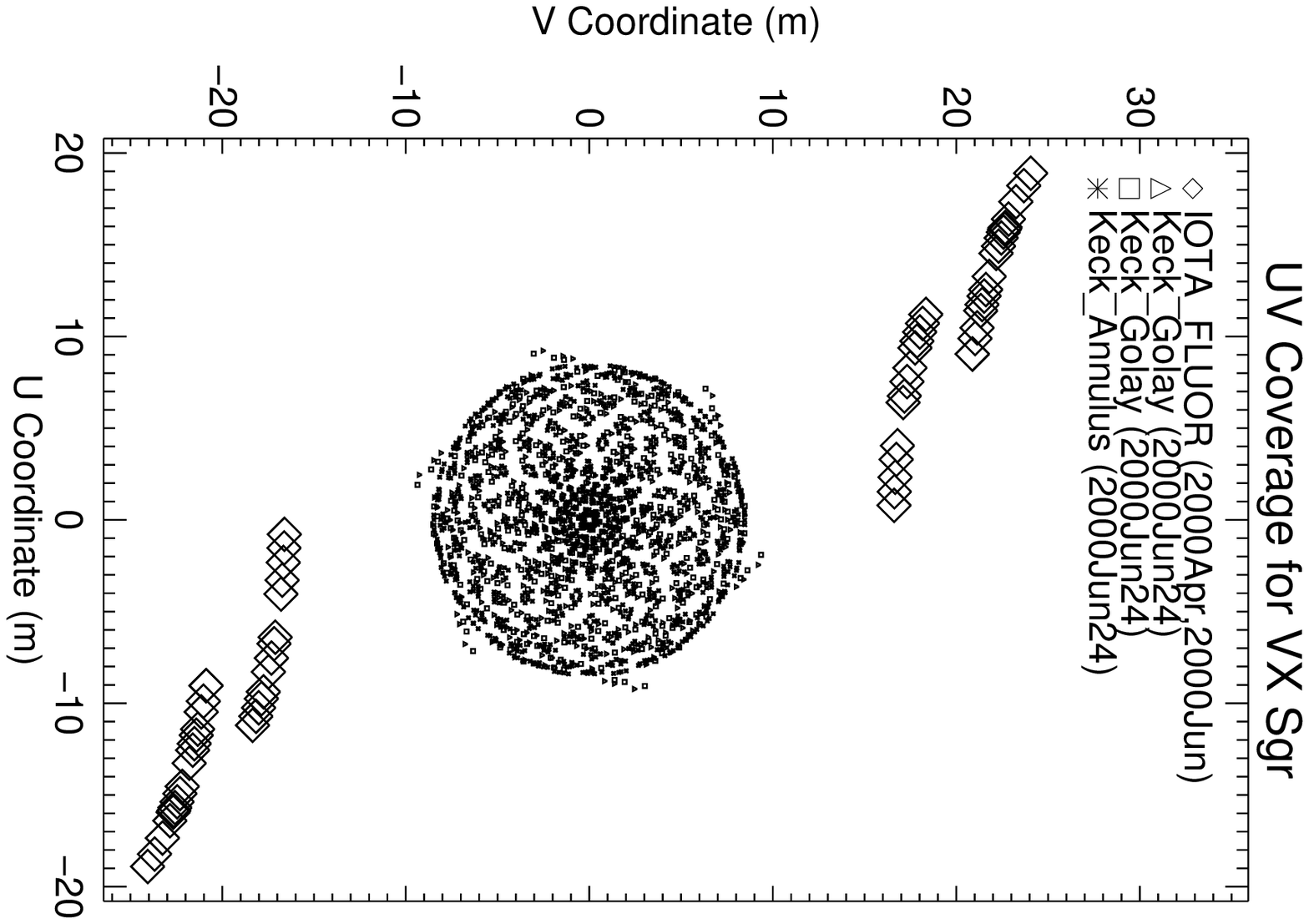}
\includegraphics[angle=90,height=3in]{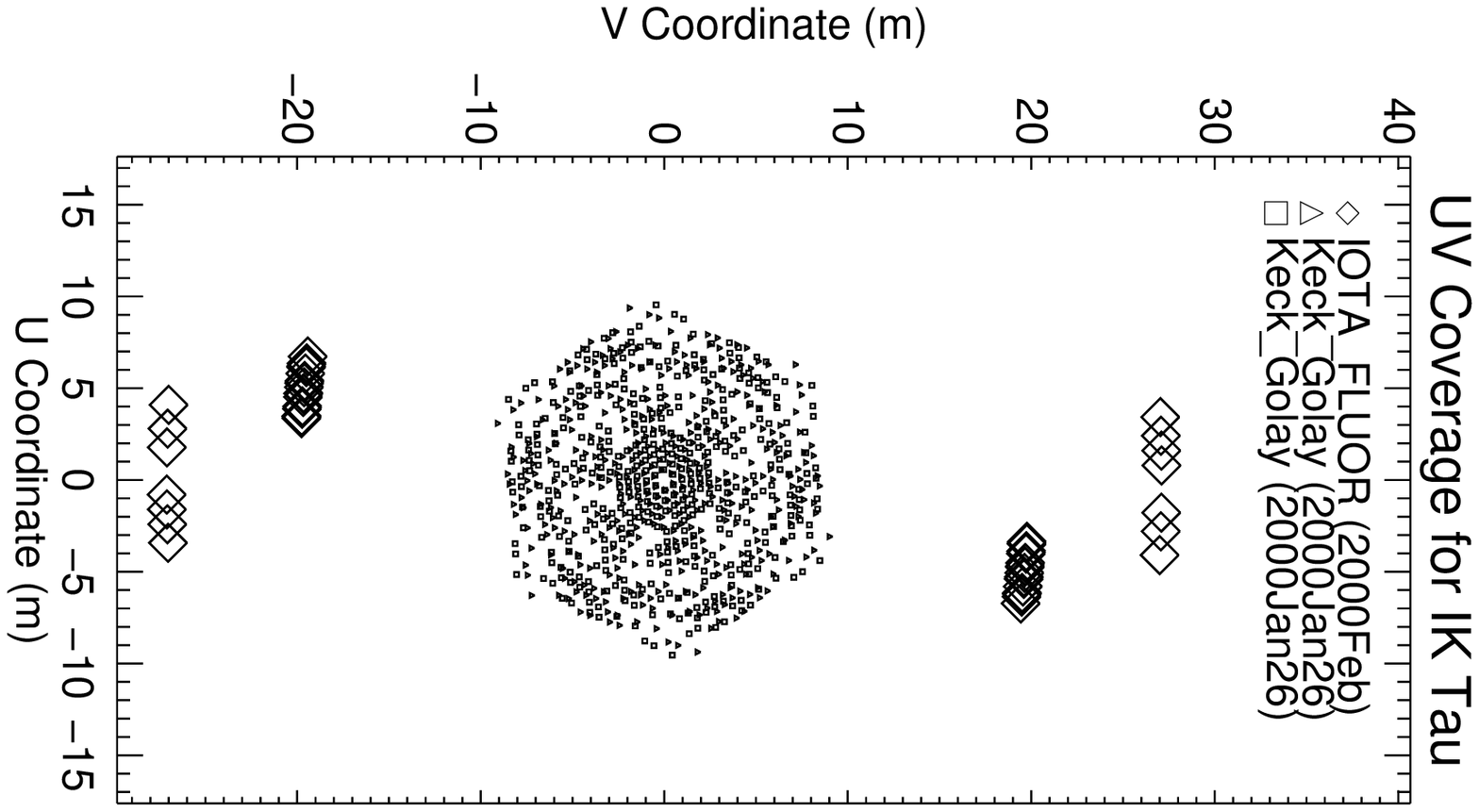}
\figcaption{UV coverage of the new interferometric observations for: a) NML~Cyg, b) VX~Sgr,
c) IK~Tau
\label{uvcov2}}
\end{center}
\end{figure}

Figure~\ref{uvcov1} \& \ref{uvcov2} contain the (u,v) coverage of our 
observations. Most targets were observed on at least two IOTA baselines, and all
have extensive visibility data for baselines shorter than 9~m from Keck aperture
masking.  The IOTA baselines tend to be oriented mostly N/S due to the 
geometry of the array.   Because of this, low declination sources have 
relatively smaller (v)-components, reducing the attainable angular resolution.

\subsection{``Peculiar'' A2I supergiant HD~62623}
\label{hd62623}
The peculiar A2I supergiant HD~62623 has a significant IR excess
recently modelled as due to the presence of circumstellar dust
\citep{plets1995, bittar2001}.  \citet{bittar2001} used multi-wavelength
Keck aperture masking data to constrain radiative transfer
models of the putative dust shell. 
Here, we present additional Keck data and 
longer baseline IOTA data which confirm that the dust shell is indeed
partially resolved.  

No definitive signs of asymmetry were found in the 2-dimensional
visibilities and closure phases of HD~62623 at 2.2$\mu$m for
3~different epochs.  If the dust in this system is arranged in a
circumbinary disk around the short-period ($\sim$137 days) binary in
this system \citep[binary separation is unresolved by the
interferometer, major axis a $\simle$ 0.1\,AU $\sim$ 0.14~mas at
700~pc,][]{plets1995}, we conclude that the disk is viewed at a
relatively low inclination (i.e., the disk is near face-on, if the
disk geometry is valid). 

\begin{figure}[thb]
\begin{center}
%\epsscale{.5}
\includegraphics[angle=90,height=3in]{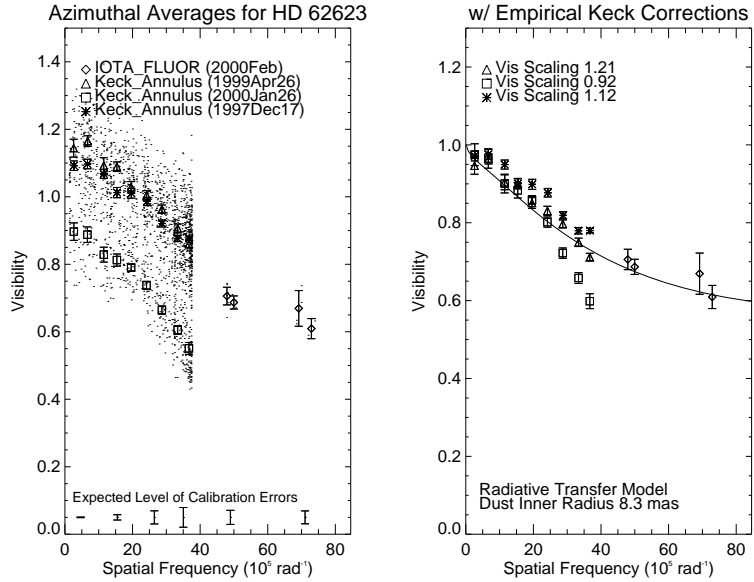}
\figcaption{HD~62623 Data. a) The left panel shows each individual Keck
and IOTA visibility datum as a point, while the azimuthal averages are
plotted with error bars.  At the bottom, an estimate of the
calibration/systematic errors are shown for each baseline range.  b)
The right panel shows only averaged data and the empirical Keck
corrections, particularly large for this source, have been applied
(see \S\ref{systematics}). The solid line represents the radiative
transfer model fit discussed in the text (\S\ref{hd62623}).
\label{hd62623c}}
\end{center}
\end{figure}

Figure~\ref{hd62623c} shows the azimuthally averaged Keck masking data
along with the longer baseline data from IOTA.  The IOTA data is
consistent with the Keck data and confirms that the dust shell is
partially resolved.  In order to further explore the consequences of
these observations, we have generated a simple radiative transfer
model that can fit the visibility data (included on
Figure~\ref{hd62623c}).

Our choice of stellar and dust shell parameters came mostly from the
previous modeling work of \citet{plets1995} and \citet{bittar2001}:
$R_\ast=0.33$\,mas, $T_\ast=9000$\,K, $R_{\rm inner}=8.3$\,mas
($T_{\rm inner}=1500$\,K),
$R_{\rm outer}=1000 R_\ast$, $\rho\propto r^{-1.3}$ \citep{plets1995},
$\tau_{2.2}=0.16$, $a=0.75\mu$m.
Note that at a distance of 700~pc, this inner radius corresponds to 5.8~AU, much
larger than the binary separation ($\simle$0.1~AU).
For this modeling we used the 
publically-available radiative transfer code DUSTY \citep{dusty}, incorporating
the (warm) silicate optical constants of \citet{ohm92}.

We emphasize that a large grain size ($a=0.75\mu$m) was used in 
order to fit the visibility data \citep[as found first
by][]{bittar2001}, otherwise the dust would be heated in excess of the
expected sublimation temperature $T\sim1500$K.  While our model fits
the visibility reasonably well, the dust does not produce enough
infrared emission to match the observed SEDs (photometry extracted
from Keck data
yielded K mag 2.32$\pm$0.10, consistent with other recent
IR photometry).  Because the shell is optically thin, it is difficult to
imagine a solution to this discrepancy, even if we abandon the
assumption of spherical symmetry.

An alternate theory for the infrared excess was explored by
\citet{rovero1994}, who found that it could be explained using only
free-free/free-bound emission from a chromosphere without any dust.
However, this model failed to explain the silicate feature seen
in IRAS-LRS spectra \citep{plets1995}. Further, the chromospheric models
predict emission arising within a few stellar radii of the photosphere
\citep[e.g.,][]{lamers1984}, so close as to be unresolved by
our interferometer observations.

The presence of near-infrared chromospheric emission, however, could
help explain the inability of dust shell models to simultaneously fit
the SED and near-IR visibility data.  This extra emission would not be
resolved but would contribute flux to the SED. 
Future observations with $\sim$10$\times$ greater
resolution, such as with the CHARA Interferometer, can directly test
this by resolving any chromospheric emission itself.

\subsection{Rapidly-evolving F supergiant IRC+10420}

IRC~+10420 is an F supergiant, surrounded by a dust shell, thought to be caught in
the short-lived phase of stellar evolution evolving from a red supergiant
into a Wolf-Rayet star \citep{oudmaijer1996,blocker1999,
humphreys2002}.  The circumstellar envelope is known to be complex on
large and small scales \citep{humphreys1997, blocker1999}.

\begin{figure}[thb]
\begin{center}
%\epsscale{.5}
\includegraphics[angle=90,height=3in]{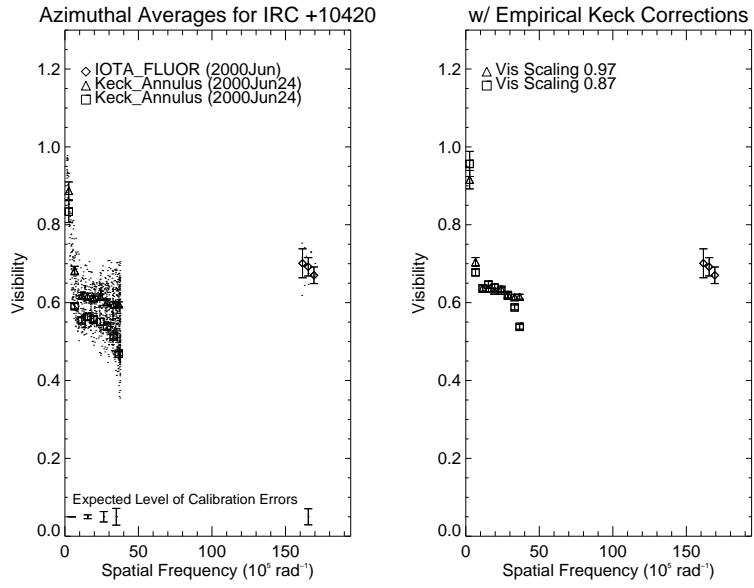}
\figcaption{Same as Figure~\ref{hd62623c}, except for IRC~+10420 data.
\label{irc10420c}}
\end{center}
\end{figure}

Figure~\ref{irc10420c} shows IOTA data and the azimuthally-averaged
Keck visibility data.  The emission from the dust shell ($\sim$38\% of
total K band flux) is resolved out on short baselines ($\simle$2m),
and then there is a visibility plateau.  As was the case for HD~62623,
there are low-level asymmetries present in the Keck data (not shown)
which we ascribe to residual miscalibration (the same asymmetry is not
present in independent observations).  Our results are consistent
with similar observations at this wavelength published by
\citet{blocker1999} for baselines $\simle$6m.  The IOTA visibilities
are slightly higher than the long-baseline Keck data, but consistent
with our expected calibration (see \S\ref{validation}).

The IOTA data extends our angular resolution of this system by a
factor of $\sim$6 from previous observations.  The fact that the
visibility appears to remain constant from 9~m out to 36~m supports a
model with an unresolved central source (diameter $<3$~mas) containing
$\sim$62\% of the K-band flux; there is no evidence for a binary
companion.  We also note that photometry extracted from the Keck
observations (K band: 3.63$\pm$0.10 mag) is consistent with the trend
of decreasing brightness documented by \citet{oudmaijer1996}.

Under normal circumstances, we would attempt to fit the visibility
data with a simple radiative transfer model to estimate physical
parameters of the dust shell.  However, for this source,
\citet{blocker1999} has convincingly shown that the short-baseline
visibility data can not be fit by a simple dust-shell model.
Furthermore, complicated dust shell features are present in HST scattered light
images by \citet{humphreys1997}, who argue
that we are viewing a bipolar outflow from a near-polar direction,
thus spherically symmetric modelling is difficult to justify for this
thick dust shell.

Modeling this system at the required level of sophistication is beyond
the scope of this paper.  We attempted to produce images of IRC~+10420 using
the aperture synthesis software (based on maximum entropy
method). Unfortunately, mis-calibration in the Keck
data corrupted the short baseline visibility (see \S\ref{systematics}).  Since
these short baselines are critical for reconstructing the large-scale
structure present in this source, imaging must await better calibrated data.

\subsection{Red Supergiant VY~CMa}
\label{vycma}
VY~CMa is a red supergiant with extreme mass loss and high luminosity
$>10^5$\lsun, approaching its end as a Type II supernova
\citep[see][and references therein, for recent summary of the
properties of this source]{monnier99a}.  While the mid-infrared
emission of the extensive dust shell around VY~CMa can be fit by a
spherically-symmetric outflow \citep{monnier2000b}, the visible and
near-infrared emission is dramatically asymmetric
\citep{kw98,monnier99a,smith2001}.  \citet{monnier99a} used Keck
aperture masking data to image the dust shell around this source at
three infrared wavelengths.  Here, we improve upon this work by
incorporating the higher resolution IOTA data.

\begin{figure}[thb]
\begin{center}
%\epsscale{.5}
\includegraphics[angle=90,width=4in]{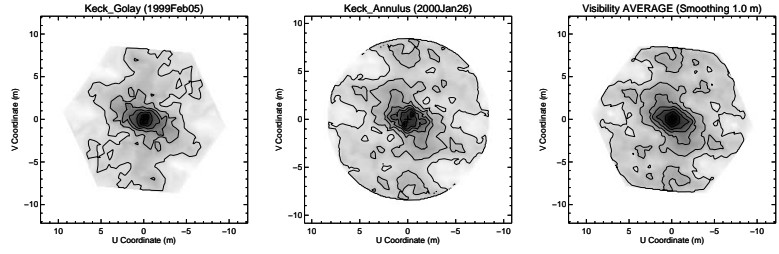}
\figcaption{Visibility data for Keck aperture  masking observations of
VY~CMa: 1999 February (left panel), 2000 January (middle panel), averaged and
smoothed (right panel).
Each solid contour line represents 0.10 in visibility
\label{vycmab}}
\end{center}
\end{figure}

\begin{figure}[thb]
\begin{center}
%\epsscale{.5}
\includegraphics[angle=90,height=3in]{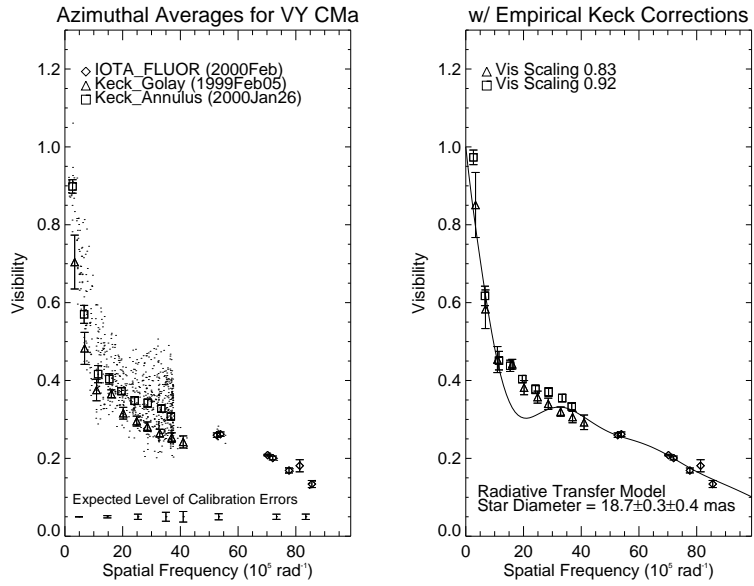}
\figcaption{Same as Figure~\ref{hd62623c}, except for VY~CMa data.  The solid
line shows the visibility prediction from a simple radiative transfer model
(see \S\ref{vycma}).
\label{vycmac}}
\end{center}
\end{figure}

Figure~\ref{vycmab} shows the 2-dimensional visibility data of VY CMa
for Keck masking observations of 1999 February and 2000 January.
Both datasets show striking asymmetric structure consistent with previous
epochs.  These data are azimuthally-averaged and included with the new
IOTA data in Figure~\ref{vycmac}.  
The IOTA data allows the stellar
component to be definitively separated from the dust component, and
further yields a direct measurement of VY~CMa's diameter.  Firstly,
we fit the diameter only using the IOTA data (result included in
Table~\ref{diameters}): 18.7$\pm$0.3$\pm$0.4 mas which contributes
$\sim$36\% of flux at 2.2$\mu$m.  This is in reasonable agreement with a
previous estimate of $\sim$20\,mas assuming a $T_{\rm eff}\sim2700$
\citep{monnier2000b,sidaner96}.  We emphasize that we measure
an apparent diameter at this wavelength, and relation to the true
``photospheric'' diameter relies on additional assumptions of
limb-darkening and other effects; late-type stars are known
to have different apparent sizes between the visible, near-IR and
mid-IR \citep{weiner2000}.

We have modelled the dust shell using a spherically-symmetric radiative
transfer model in order to illustrate how one can be misled by 
models when 
the true source structure is asymmetric and complex.
The right-panel of Figure~\ref{vycmac} shows the visibility curve of a
model with the following parameters: $R_\ast=9.35$\,mas,
$T_\ast=2600$K, $R_{\rm inner}=65$\,mas ($T_{\rm inner}=1050K$),
$R_{\rm outer}=5000$\,mas, $\rho\propto r^{-2}$ (uniform outflow),
$\tau_{2.2}=2.18$.  The effective temperature was found by fitting to
the K band magnitude of VY~CMa on 2000Jan26 of +0.1$\pm$0.1, based on
photometry extracted from Keck data itself. 
Since we are using a simple blackbody function to
estimate the stellar flux and because we are not fitting to the total
luminosity, this temperature is not  definitive
(although the diameter measurement is direct).  Assuming a distance of
1.5~kpc \citep{monnier99a}, the above angular quantities correspond to
$R_\ast=14$\,AU and $R_{\rm inner}=97.5$\,AU.

For this source, and the ones that follow, we have used the 
Wolfire radiative transfer code \citet{wolfire86} instead of DUSTY,
which does not handle cases when the dust temperature is more than
$\frac{1}{2}$ the photospheric temperature, as is often appropriate for the
most evolved red giants and supergiants.  Here we used warm silicate
optical constants of \citet{ohm92}, with Mie scattering calculations
based on \citet{toon81}, assuming MRN grain size distribution
\citep{mrn}. Details on use of this code have been given previously by
\citet{danchi1994} and  \citet{monnier1997}.

\begin{figure}[thb]
\begin{center}   
%\epsscale{.5}
  \includegraphics[angle=0,height=3in]{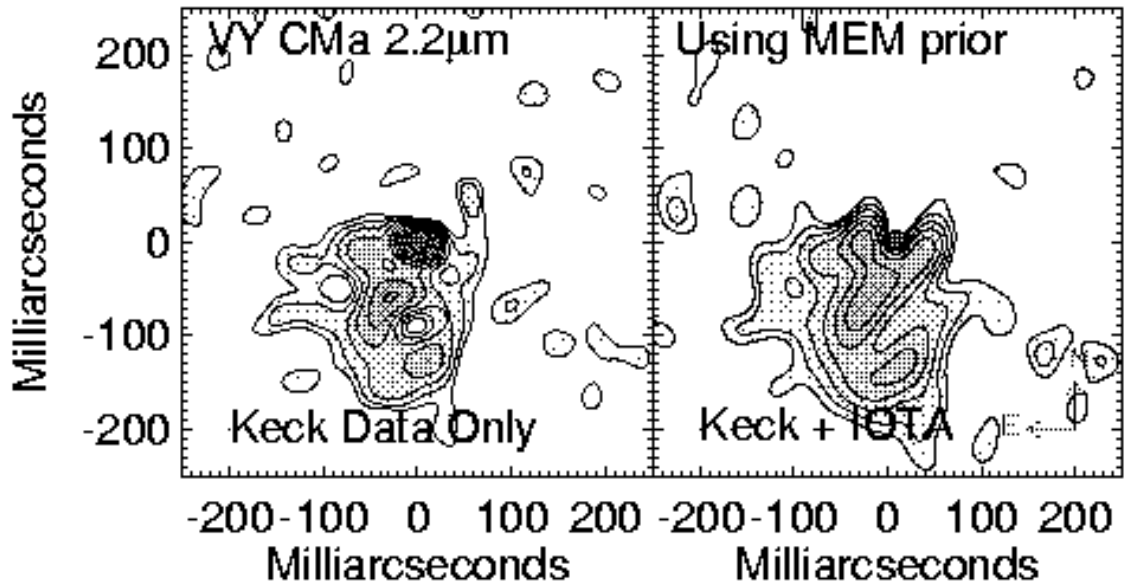}
\figcaption{Maximum entropy image reconstructions of the VY~CMa circumstellar
  environment. The left panel shows an image reconstruction using Keck
  masking data only and a uniform prior.  The right panel shows an
  image reconstruction using a MEM prior incorporating a 18 mas disk
  in the center of the asymmetric nebula (see text for further
  details); the star is shown here actual size.  The lowest contour
  level in each figure represents a 2-$\sigma$ noise level above the
  background; the subsequent contours are logrithmically-spaced,
  increasing by a factor of 2 for each level.  For reference, the
  1-$\sigma$ noise limits are 0.09\% and 0.026\% of the peak for the
  left and right panel respectively, a scaling due to the difference in the
  compactness of the central source.
\label{vycma_images}}
\end{center}
\end{figure}

As can be seen in Figure~\ref{vycmac}, the fit to the visibility data
is reasonable at short and long baselines, but poor at intermediate
scales.  This fit could be improved by changing the assumed dust properties or
adding another dust shell in order to modify the visibility curve.
However in this case, unmistakably evidence in the closure phases and
visibility amplitudes show that
the deviation here comes from the fact that
the dust shell is highly asymmetric.  To better visualize the dust
distribution, we have incorporated the high-resolution IOTA data into
the image reconstruction process, although current software
limitations (see \S\ref{masking}) required an {\em ad hoc}
approach. Figure~\ref{vycma_images} shows the image reconstruction
results using the 1999 February masking data, with and without IOTA
information.  The left panel shows the image using only Keck data, and
this epoch looks very similar to previous ones published by
\citet{monnier99a}.

When using Keck data only, the central source appears here to be
slightly resolved and elongated (see left panel of
Figure~\ref{vycma_images}).  While it is not impossible that the
central star of VY~CMa is highly elongated, it is more likely an
artifact of the MEM algorithm \citep{nn86} which attempts to spread
out the light as much as possible consistent with the maximum spatial
resolution of the data.  In Figure~\ref{vycmac} which incorporates
IOTA data, we have shown that
the central star contributes $\sim$36\% of the K-band flux and is
$\sim$18~mas in size.  We can include this high-resolution information
in the MEM
fit by using the MEM prior, which is the default map that MEM uses 
when the data cannot constrain the solution.
The technique of using the MEM prior to
incorporate the presence of a compact central source known from either the
SED or longer-baseline data, has already been explored by
\citet{monnier_spie2003a} and \citet{lkha2}; more discussion of this method 
can be found in these references.

The right panel of Figure~\ref{vycma_images} shows the image
reconstruction using a prior of an 18~mas star surrounded by
asymmetric extended emission \citep[based on previous imaging
of][]{monnier99a}.  While these two image reconstructions are very
similar to each other (indeed, both fit the data with a reduced
$\chi^2\sim$1), there are some details in the new image which are
important.  The dust distribution forms more of an arc to the south of
the star and is less 'clumpy.'  Without sufficient long-baseline (u,v)
coverage, the MEM (or any other) method by itself can not precisely
image dust very close to the stellar photosphere without the
additional information of the stellar size and flux contribution.

Interpretation of this dust shell is complicated by its high optical depth.
These new results confirm previous indications by \citet{kw98} and
\citet{monnier99a} of bipolar dust distribution
(the dusty ``disk'' is oriented roughly E/W). The K-band light
arises predominantly from the southern ``pole'' of the dust envelope,
where the relatively low optical depth allows hot dust emission near
the star to be seen directly and also allows scattered light to escape
into our line-of-sight.  With high-fidelity images of this complicated
dust shell, we can begin proper motions studies, as has already been
demonstrated for IRC~+10216 \citep{tuthill2000a}.  We hope to image new dust
production episodes as they happen and to deduce the physics of mass-loss by
following the time evolution of the circumstellar environment.

\subsection{Red Supergiant NML~Cyg}

NML~Cyg is an extreme red supergiant surrounded by an
optically-thick dust envelope.  Mid-infrared interferometry uncovered
strong evidence for multiple shells of dust \citep{monnier1997}.  This
basic result was confirmed and explored further by new near-infrared
speckle measurements of \citet{blocker2001}.  Our new observations
allow this dust shell to be imaged with unprecedented fidelity
by separating the dust emssion from the stellar emission.

\begin{figure}[thb]
\begin{center}
%\epsscale{.5}
\includegraphics[angle=90,width=4in]{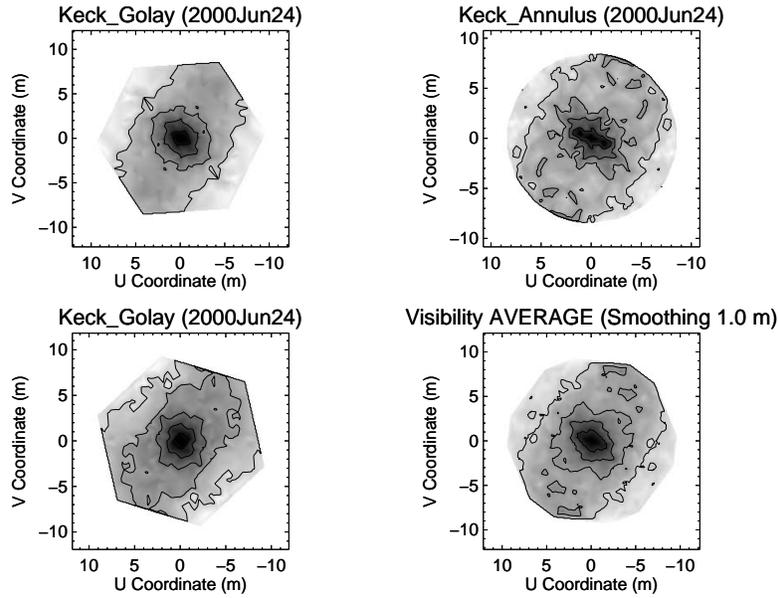}
\figcaption{Visibility data for Keck aperture masking observations of
NML~Cyg (three measurements from 2000Jun); the bottom-right panel is
the average of the other three, and has been slightly smoothed. Each
solid contour line represents 0.10 in visibility. The NE-SW elongation
of the source (in visibility space, the source is more resolved along
the NE-SW axis) is real and reflects a bipolar dust distribution
imaged in Figure~\ref{nmlcyg_images}.
\label{nmlcygb}}
\end{center}
\end{figure}

Figure~\ref{nmlcygb} shows three separate Keck masking observations of
NML~Cyg.  As was the case for VY~CMa, the strong asymmetry is repeated
in each independent measurement and thus can be reliably associated
with source structure and not miscalibration.  NML~Cyg does not show
large closure phases, indicating the emission is largely
centro-symmetric \citep{monnier_mss}, in marked contrast to VY~CMa
which showed large closure phases resulting from the highly asymmetric
nebula (see Figure~\ref{vycma_images}).  Photometry from Keck found
NML~Cyg Kmag +0.55$\pm$0.10 at this epoch.

\begin{figure}[thb]
\begin{center}
%\epsscale{.5}
\includegraphics[angle=90,height=3in]{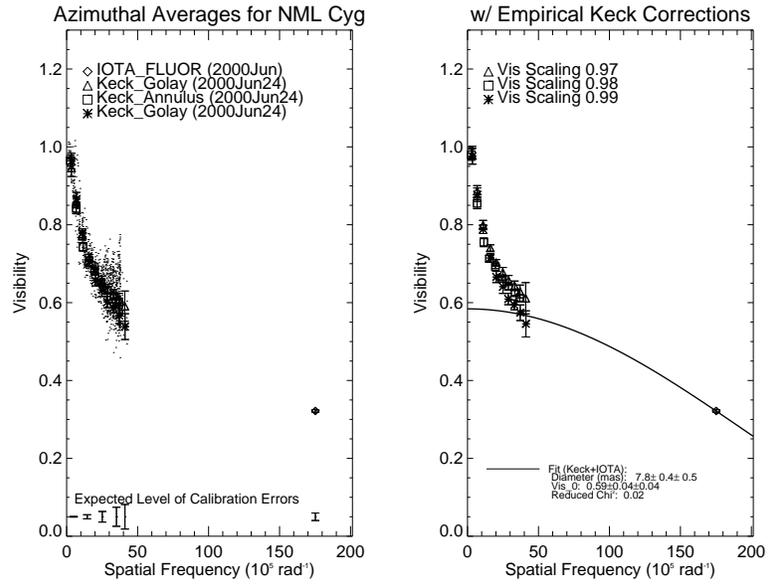}
\figcaption{Same as Figure~\ref{hd62623c}, except for NML~Cyg data.
The solid line shows a uniform disk fit to the longest baseline visibility
data.
\label{nmlcygc}}
\end{center}
\end{figure}

Figure~\ref{nmlcygc} shows the azimuthally averaged Keck data along
with limited IOTA measurements.  The large gap between the two
baseline ranges make interpolation uncertain.  A uniform disk was fit
to the longest-baseline Keck data and IOTA visibilities, and the
result (diameter 7.8$\pm$0.4$\pm$0.5~mas) is shown in the right panel
of this figure; this diameter estimate should be considered
an upper limit until more extensive data fully characterizes the 
``knee'' or ``break'' in the visibility
curve between circumstellar and photospheric emission.
 Since the bolometric luminosity is well-constrained
by the SED \citep[$\sim$3.5~10$^5$\,\lsun at 1.8kpc,][]{monnier1997},
this diameter implies T$_{\rm eff}\sim$3650~K, somewhat hotter than
expected for an M6 supergiant \citep[e.g., $\sim$3375~K,][]{vanbelle1999}.

Given the recent extensive efforts to model this dust shell, an
overly-simplistic treatment here would serve little purpose. Instead,
we present aperture synthesis images of the dust shell which allow us
to discover asymmetries and clumpiness in a model-independent way.

\begin{figure}[thb]
\begin{center}
%\epsscale{.5}
\includegraphics[angle=0,height=3in]{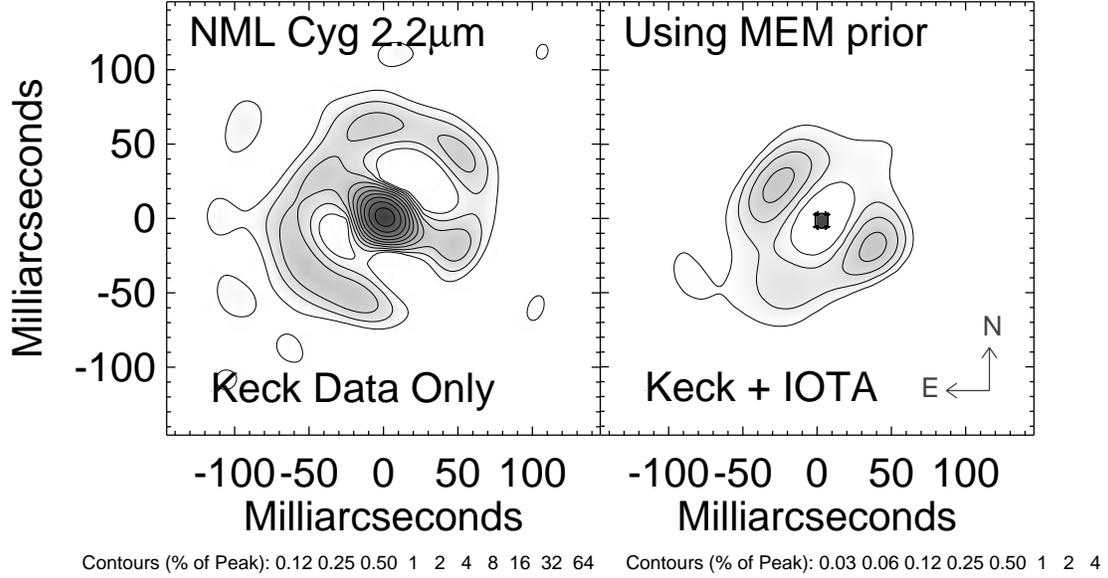}
\figcaption{Maximum entropy image reconstructions of the NML~Cyg
circumstellar environment. The left panel shows image reconstruction
Keck masking data only, using a uniform prior. 
The right panel shows an image reconstruction using a MEM prior with
59\% of the flux in a single 7\,mas pixel (the star is shown here
actual size).  This image reconstruction
allows a high fidelity dust shell image to be created by constraining
the size and amount of compact stellar emission (based on IOTA data).
The logarithmic contour levels each
represent a factor of 2 in surface brightness compared to the peak.
for the left panel, we have 0.125, 0.25, 0.5, 1, 2, 4, 8, 16, 32, and
64\% respectively; for the right panel, we have 0.03125, 0.0625, 0.125,
0.25, 0.5, 1, 2, and 4\% respectively.
\label{nmlcyg_images}}
\end{center}
\end{figure}

We present image reconstructions following the strategy adopted in the
last subsection for VY~CMa. Figure~\ref{nmlcyg_images} shows both
image reconstruction with Keck masking data alone (left panel) and
using a MEM prior to introduce the presence of an unresolved central
source (with 59\% of flux, based on IOTA data).  In this case (unlike
VY CMa earlier) the additional prior information has made a dramatic
difference between these images.  In the left panel, without the MEM
prior containing the central source, the algorithm has created an
image with a very elongated central source to fit the asymmetric
visibility data. The size and shape of this central source is not
physically plausible (i.e, a red supergiant photosphere is not
expected to be this large and elongated), and thus we use a MEM prior
to incorporate {\em a priori} information (derived from IOTA data)
concerning the size and shape of the central source.

This example serves as a potent reminder that MEM imaging of extended
dust shells around unresolved ``point'' sources depends greatly on the
resolution of the interferometer {\em in the case that the dust shell
is marginally resolved}.  In this case, there is simply not enough
information in the Keck data alone to constrain the large number of
images consistent with the visibility data and closure phases.
Indeed, both of these images fit the Keck data with a reduced
$\chi^2\sim$1; it is the addition of {\em a priori} information
regarding the nature of the stellar component that allows a higher
fidelity image reconstruction.  

The new image significantly advances our understanding of the NML~Cyg
dust shell by establishing that the inner circumstellar shell is not
spherically symmetric.  Astrophysically, the ``Keck + IOTA'' image can
be understood in the context of the H$_2$O maser data of
\citet{richards96}, who found evidence already for a bipolar outflow
along the NW/SE axis.  We interpret our data as the first definitive
detection of the dust asymmetry, showing an equatorial enhancement
along the NE-SW axis, perpendicular to the 
maser outflow.  This
identification could not have been made without combining the Keck
with the IOTA data.  The SiO maser data of \citet{boboltz2000} has
been interpreted as a sign of rotation about a NW-SE axis: it is
interesting to speculate that the increased dust density seen in our
image may be due to stellar rotation.

Bipolar outflows and the associated dust shell asymmetries are not
reliably modelled with a spherically-symmetric radiative transfer code
when the dust shell is optically thick.  This can help explain the
strong difficulty in fitting the near-IR visibility data at the same
time as the SED and mid-IR visibility data
\citep{monnier1997,blocker2001}.  The near-IR visibility is strongly
affected by the fact that the average dust shell optical depth (which
controls the total near-IR emission) is different to the line-of-sight
optical depth (which affects the stellar contribution due to
extinction).  Assuming a different stellar fraction ($\neq59$\%) causes
the reconstructed dust shell to change somewhat in scale but not
general morphology;
additional data with resolution intermediate between
Keck and this IOTA data will allow both the stellar diameter and
fractional flux to be precisely measured.
Until then, current conclusions should remain qualitative.

\subsection{Red Supergiant VX Sgr}
\label{vxsgr}
VX~Sgr is a bright infrared source with strong maser emission, a red
supergiant experiencing heavy mass loss.
There has been no high-resolution near-IR data published since early
speckle results of \citet{dyck84}, when the dust shell
was only partially resolved.
Our new data has $\sim$5$\times$ greater resolution and the dust shell
is easily resolved at short baselines ($\sim$3~m), allowing the dust
and stellar components to be distinguished even without long baseline
IOTA data.  The two-dimensional visibility data from Keck show some
evidence for asymmetry (i.e., deviations from circularity); however,
since the dust contributes only $\sim$20\% of the K-band flux, we can
not place strong limits on possible asymmetries, given our calibration
uncertainties. As was the case for NML~Cyg previously, the closure
phases for VX~Sgr are all small ($\simle$3 degrees), indicating the
dust shell is centro-symmetric.

\begin{figure}[thb]
\begin{center}
%\epsscale{.5}
\includegraphics[angle=90,height=3in]{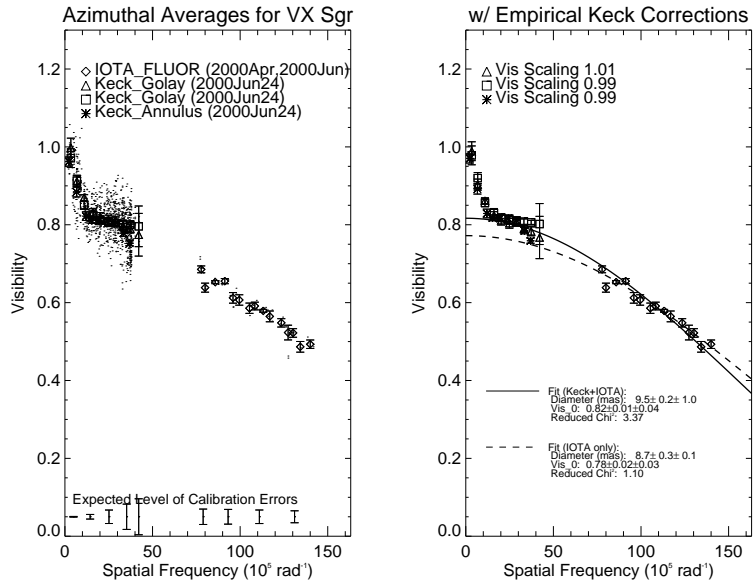}
\figcaption{Same as Figure~\ref{hd62623c}, except for VX~Sgr data.
The solid and dashed lines show two different Uniform Disk fits to
the stellar photosphere (see \S\ref{vxsgr} for further information).
\label{vxsgrc}}
\end{center}
\end{figure}

Figure~\ref{vxsgrc} shows the azimuthal-averages of the Keck data,
along with extensive IOTA data allowing a diameter measurement of
the underlying star. 
The most precise measurement comes by fitting to
the IOTA data alone, which has sufficient baseline coverage to
constrain the diameter 8.7$\pm$0.3$\pm$0.1 mas; this fit can be found
in the figure.  Also shown is a fit which includes the longest
baseline Keck data: 9.5$\pm$0.2$\pm$1.0 mas.  The latter estimate has
a large systematic error due to known systematic errors in the Keck
calibration at long baselines, however the two fits are statistically
consistent.  We conclude there is a 5\% calibration difference between
the IOTA and Keck datasets, although we can not determine the cause
(e.g., atmospheric miscalibration, filter bandpass differences).

\begin{figure}[thb]
\begin{center}
%\epsscale{.5}
\includegraphics[angle=90,height=3in]{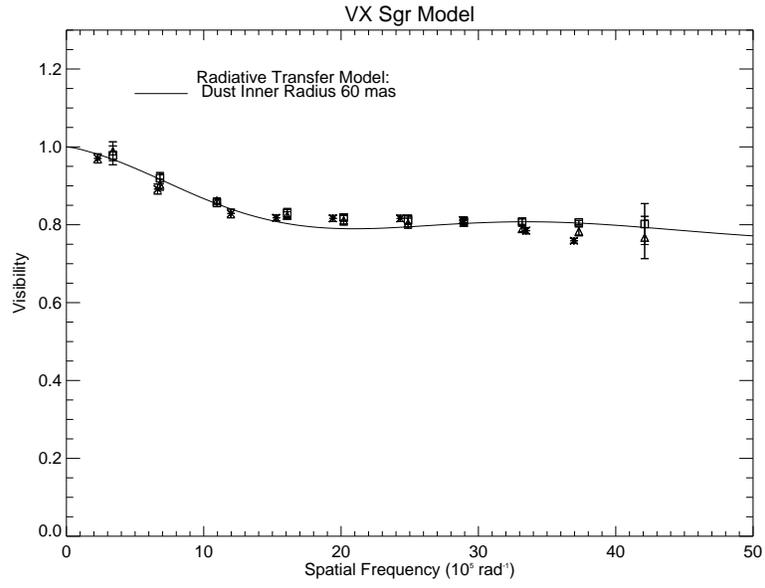}
\figcaption{This figure shows the radiative transfer fit to the VX~Sgr
2.2$\mu$m visibility data. 
\label{vxsgr_models}}
\end{center}
\end{figure}

Despite the calibration difficulties with the aperture
masking which hampers measurements of asymmetries, the azimuthal
averages of the three different masking datasets shown in Figure~\ref{vxsgrc}
are quite consistent with each other and motivates us to pursue radiative
transfer modeling.
Because the Keck data resolves the dust shell completely, we have
performed modelling using this data set alone (without IOTA data explicitly, but using the
angular diameter derived above), also incorporating the
results of Keck photometry, 
K-band -0.2$\pm$0.1 mag.  
Figure~\ref{vxsgr_models} shows the reasonable fit for a simple model:
$R_\ast=4.35$\,mas, $T_\ast=3200$K, $R_{\rm inner}=60$\,mas ($T_{\rm
inner}=940K$), $R_{\rm outer}=5000$\,mas, $\rho\propto r^{-2}$
(uniform outflow), $\tau_{2.2}=0.17$, and with dust properties
and grain size distribution the same as described previously for VY~CMa.

The diameter of VX~Sgr found here is dramatically smaller than assumed
in a recent modeling paper of \citet{greenhill95}.  In this paper,
mid-IR interferometry from the ISI was used to constrain a radiative
transfer model with a stellar diameter of 26~mas (3$\times$
larger than found here!).  It is not clear why this previous model
assumed such a large photosphere (and correspondingly low effective
temperature) since the 10 micron visibility data did not have enough
resolution to directly constrain the diameter as we do here.  The SiO
masers at $\sim$16.9~mas can now be interpreted to lie at 3.9
R$_\ast$ (instead of 1.3 R$_\ast$) -- a significant difference, 
showing that SiO masers do form well above the photosphere.  We note
that the dust shell parameters (which {\em were} constrained by the ISI
mid-IR measurements) of \citet{greenhill95} are quite consistent with
our current modeling of the near-IR Keck visibility data.

VX~Sgr is a good source for future study, since the dust shell is
fairly large and the high SNR closure phases from Keck show it to be
centro-symmetric (recall that disk structures possess centro-symmetry,
thus closure phases can not help in distinguishing circular dust shells
from disks).  In \S\ref{imaging}, we discuss some lessons learned
regarding imaging dust shells around bright sources, 
such as VX~Sgr.

\subsection{O-rich Mira IK~Tau}
\label{iktau}
IK~Tau is an evolved Mira variable star with an optically-thick,
silicate-rich dust envelope.  Here, we report the first
high-resolution near-IR results since \citet{dyck84}, extending full
(u,v) coverage by a factor of $\sim$3.
The Keck masking closure phases are small (close to zero), thus the
dust shell appears centro-symmetric at this resolution
(scales$\simle$50~mas). As for VX~Sgr discussed earlier, the 2D
visibility data from masking shows signs of asymmetries which 
could not be confirmed due to poor data quality.
Future observations will focus on the 2D visibilities, while we
consider here only the azimuthally-averaged visibility in the context
of model-fitting.

\begin{figure}[thb]
\begin{center}
%\epsscale{.5}
\includegraphics[angle=90,height=3in]{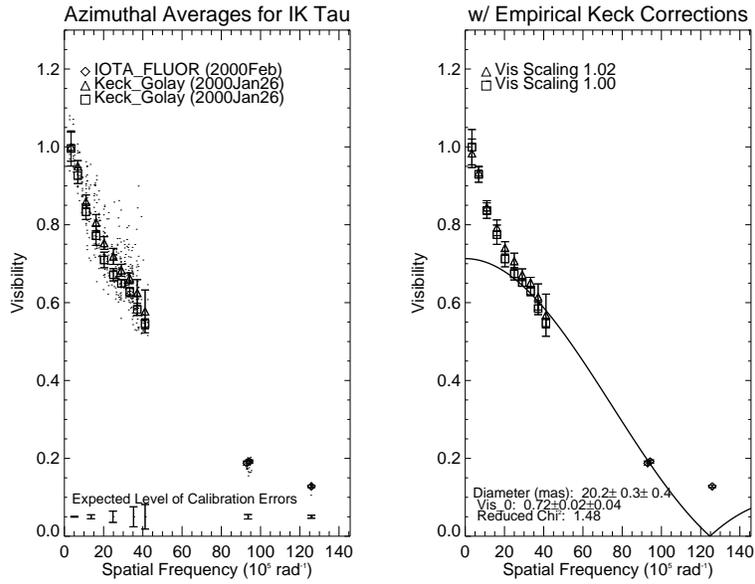}
\figcaption{Same as Figure~\ref{hd62623c}, except for IK~Tau data.
The solid line shows a uniform disk fit which ignores the
visibility datum at $\sim$27~m baseline 
(see text \S\ref{iktau} for justification).
\label{iktauc}}
\end{center}
\end{figure}

Figure~\ref{iktauc} shows the azimuthally-averaged Keck data along
with longer-baseline IOTA data.  This combination allows the diameter
of IK~Tau to be measured for the first time. However, the large gap in
baseline coverage between the longest Keck baselines and the IOTA
baselines leaves some ambiguity for what data to use.  One possibility
is to fit a uniform disk to the IOTA data alone, which results in a
diameter of 12.4$\pm$0.4$\pm$0.1~mas.  There are two major problems
with this result.  First, this small size would require an effective
temperature $\simge3000~K$ in order to have sufficient luminosity to
match observed flux (based on Keck photometry, the K band magnitude of
IK~Tau on 2000Jan26 was -1.05$\pm$0.1).  This is unlikely considering
the strong CO and H$_2$0 bands in the near-IR spectrum
\citep{hyland1972}, which suggests an $T_{\rm eff}\sim2000K$
appropriate for a star with spectral type M10III.  Furthermore, a
simple radiative transfer model fit (not shown) with this small
stellar component requires a dust shell inner radius of $R_{\rm
in}\sim10$\,mas, so close to the star that the dust would be heated to
the unrealistically-high temperature of 2300~K.

Instead, we base our uniform disk fit on the baselines longer than 7~m
and shorter than 21~m (ignoring the longest-baseline IOTA data at
27~m); the result of this fit appears in Figure~\ref{iktauc}.  The
fitted diameter, 20.2$\pm$0.2$\pm$0.3~mas, is more consistent
with expectations, corresponding to an effective temperature of
$\sim$2300\,K.  One major difficulty with this fit is that the
prediction at the longest IOTA baseline ($\sim$27\,m) is not
consistent with the measured data.  Having dismissed the small
12.4~mas diameter that would be needed to fit both sets of IOTA data,
we are left to explain this major discrepancy.

Our preferred explanation for the high visibility at the longest IOTA
baselines is that the IK~Tau photosphere has strong departures from
uniform brightness, either due to hotspots
\citep[e.g.,][]{tuthill1999} or extended molecular emission
\citep[e.g.,][]{tsuji1997,matsuura2002, jacob2002}.   
Similar effects have already been seen around other late-type O-rich Miras 
\citep{perrin1999,thompson2002a}.  Given the extreme molecular band
structures of M10III stars, this explanation takes on greater credence.
Longer baseline data (preferable with closure phases) will be required
to confirm this and to also rule out the presence of a binary companion.

\begin{figure}[thb]
\begin{center}
%\epsscale{.5}
\includegraphics[angle=90,height=3in]{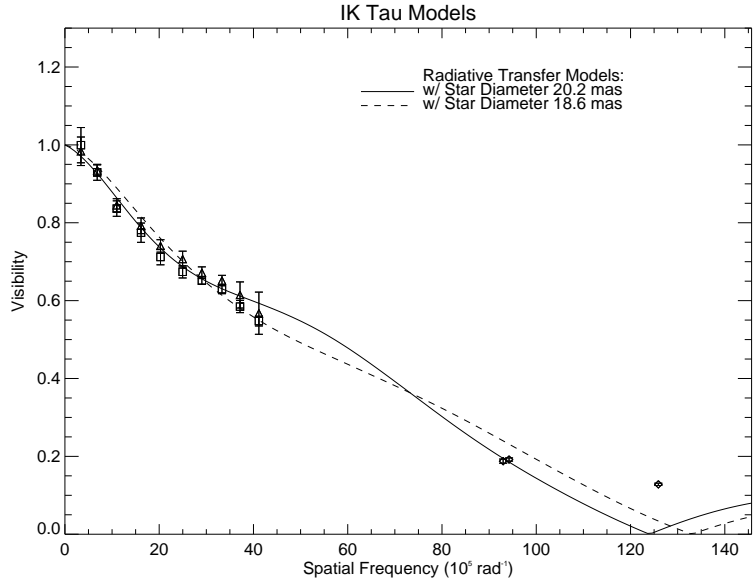}
\figcaption{This figure shows results of two fits to the IK~Tau 2.2$\mu$m
visibility data using simple radiative transfer models.  The fit to
the short baseline data is noticeably better using a stellar diameter
of 20.2\,mas than 18.6~mas; diameters smaller than this range require
dust temperature $\simge$1500~K.  The disagreement between models and
the data at the longest baselines may be due to deviations from
uniform brightness across the photosphere of this highly evolved giant
star (M10III).
\label{iktau_model}}
\end{center}
\end{figure}

A radiative transfer model has been fit to the visibility data using
the 20.2~mas photospheric diameter, and satisfactory results were
obtained.  Figure~\ref{iktau_model} shows the predicted visibility
curve at 2.2$\mu$m for a model with following parameters:
$R_\ast=10.1$\,mas, $T_\ast=2300$K, $R_{\rm inner}=35$\,mas ($T_{\rm
inner}=1100K$), $R_{\rm outer}=5000$\,mas, $\rho\propto r^{-2}$
(uniform outflow), $\tau_{2.2}=0.32$.  We used the same dust
properties as described previously for VY~CMa.  Assuming a distance
200\,pc \citep{sidaner96}, then $R_\ast = 2.0$\,AU and $R_{\rm inner}
= 7$\,AU.

The longest baseline Keck data may not be fully resolving the dust
shell, resulting in ambiguity over the fractional flux of the dust shell
relative to the star.  In order to explore the effects of this,
we generated models with successively smaller stellar diameters. 
In order to maintain a reasonable fit, decreasing
the stellar size requires increasing the dust contribution and decreasing
the dust shell inner radius.  Eventually, the inner radius becomes so
small that an unphysical dust temperature is reached.  In
Figure~\ref{iktau_model}, we have included the visibility curve for
the most extreme model with plausible dust temperatures ($T_{\rm
inner}=1500K$): $R_\ast=9.3$\,mas, $T_\ast=2500$K, $R_{\rm
inner}=22$\,mas, $\tau_{2.2}=0.27$ (other parameters the same).  This gives
a marginally poorer fit to the Keck data (and short baseline IOTA data), 
supporting
the previous model with the
larger stellar diameter and cooler dust.

As for VX~Sgr previously, we have not included a MEM image reconstruction here,
due to the uncertainty in the stellar contribution and the limited
observing set.
Future observations with better calibrated visibilities and more
uniform baseline coverage should allow high-fidelity imagery.

\subsection{General comments on imaging with optical interferometers}
\label{imaging}
In this paper, we successfully imaged only 2 of the 6 target stars
with extended dust shells, VY~CMa and NML~Cyg; a lower success rate
than anticipated.  The imaging failures resulted for a variety of
reasons, and we now discuss these in order to help other workers avoid
the same pitfalls.  While imaging work is only just beginning with
optical interferometers, the proliferation of ``imaging'' arrays with
three or more telescopes (COAST, NPOI, IOTA, ISI, CHARA, Keck, VLTI)
presages impending expansion of imaging experiments; our experiences
should prove instructive.

Images were made very early on with the Keck aperture masking
experiment for the red supergiant VY~CMa \citep{monnier99a}, dusty
pinwheel nebulae around Wolf-Rayet stars
\citep{tuthill99b,monnier1999}, carbon stars IRC~+10216 and CIT~6
\citep{tuthill2000a,monnier_cit6}, and young stars LkH$\alpha$~101 and
MWC~349 \citep{tuthill2000a,danchi2001}.  Despite suffering from the
same miscalibration problems encountered in this paper, imaging these
sources was rather straightforward.  Two major differences between
these previously published sources and the sources presented here
account for the differing ease of imaging: a) the dust shell dominated
the flux (central source contributed little flux), and b) most of the
previous dust shells were very asymmetric, possessing large non-zero
closure phases.  We now discuss the importance of each of these
characteristics for imaging.

Consider a highly resolved dust shell with little or no contribution
from an unresolved central source: the Visibility might be (say) 8\%
at some long baseline.  Mis-calibrations are typically multiplicative and
hence a 10\% error corresponds to $\Delta V$=0.008 at this baseline, a
small fraction of the total dust shell contribution (fraction
$\sim$1.0).  However, consider the case when the dust shell only
contributes 20\% of the flux (as was the case for VX~Sgr in this
paper).  This means that long-baseline visibility data (when the dust
is mostly resolved and only the visibility from the central unresolved
star is left) will be quite high, V$\sim$80\%. Hence, a 10\% error
translates to $\Delta V=0.08$, quite significant effect considering
the dust shell signal is at most only $\Delta V_{\rm shell}=0.20$.  Since the
model for the central (point) source is well-known, this has the effect of
essentially transferring {\em all} of the visibility error onto the
remaining component -- the dust shell.  

From the examples above, you can see that for the same size dust
shell, the signal-to-noise ratio of the dust shell visibilities
($\frac{\Delta V_{\rm shell}}{\Delta V_{\rm error}}$) go from
$\frac{1.0}{0.008}\sim$125 with no point source contribution to only
$\frac{0.20}{0.08}\sim$2.5 with an 80\% point source, considering just
multiplicative miscalibrations.  Miscalibrations thus have a
compounding effect when the central star dominates the flux (both the
absolute level of miscalibration increases and the proportional effect
compared to the dust shell signal increases).  We remark that Albert
Michelson, in the first interferometry experiments at Mt. Wilson
\citep{michelson21}, cleverly measured stellar diameters by finding
the visibility null (V$=$0), which is zero no matter what the
visibility miscalibration might be!

The second reason that imaging was easier with previous sources is
because many are very asymmetric.  The strong deviations from
centro-symmetry meant that much of the morphology information was
encoded in the Fourier phases and not just the Visibility amplitudes.
This effect is enhanced for sources with strong central sources which
dilute the closure phase signal of the dust shell.  As discussed
earlier in this work, the Keck masking experiment (and most
interferometers) can measure closure phases quite accurately because
atmospheric changes do not bias the measurement
\citep[e.g.,][]{monnier_mss}.	In general, any image reconstruction
procedure that incorporates
a $\chi^2$-type statistic to measure goodness-of-fit is most
constrained by high signal-to-noise ratio (SNR) data. Hence
the algorithm will implicitly rely heavily on the high SNR
phase information when making images, and would be more immune to the
miscalibrations in the visibility amplitudes.  This is the main reason
why imaging of VY~CMa (Figure~\ref{vycma_images}) showed fewer changes
than NML~Cyg (Figure~\ref{nmlcyg_images}) when long-baseline IOTA data
was incorporated.

While the above problems affect optical interferometers more than
radio interfometers, a third difficulty encountered for a few sources
here is common to all interferometers: the Fourier coverage of the
interferometer must match the angular size of the source being
observed.  For IRC~+10420, the dust shell is nearly too large for the
Keck masking experiment (over-resolved on short baselines); for
HD~62623, the baselines were not long enough  to fully-resolve the
dust shell structures.

In summary, imaging faint centro-symmetric dust shells around bright
stars is difficult for reasons both obvious and subtle.  {\em All of
the effects} described above contributed to the problems encountered
in this paper.  The dust shells for most of the sources presented here
contributed $<$50\% of the flux and showed closure phase with only
small departures from zero (the only major exception was VY~CMa).
Imaging these sources will remain challenging until excellent Fourier
coverage and excellent visibility calibration ($\sim$3\% error) can be
achieved at the same time.

An interesting consequence of the above difficulties is that there is
a tendency to successfully image ``strange-looking'' dust shells,
(e.g., VY~CMa, CIT~6, IRC+10216), but to fail to easily image
circularly-symmetric ones.  This is consistent with the fact that the
successfully-imaged sources (thus far) are not ``normal'' mass-losing
stars, but rather are extreme cases that were most suitable for early
interferometric imaging.  Although current evidence suggests that large-scale
dust shell asymmetries are common to mass-losing stars, 
too few dust shells have been imaged to say this with confidence.
As the spatial resolution and sensitivity of
interferometers improves, we should be able to image more ``normal''
evolved stars and begin to know whether strange dust shells are the
exception or the rule.

\section{Conclusions}

\label{conclusions}

Major results here fall into two categories: stellar diameters and
circumstellar dust shells.

We were able to measure 2.2$\mu$m stellar sizes of a number of
dust-obscured sources for the first time.  The VX~Sgr diameter was
found to be about 3$\times$ smaller than previous modeling, with
important repercussions for understanding the SiO maser distribution.
The IK~Tau and NML~Cyg data suggest either photospheric profiles that
strongly deviate from a uniform disk or the presence of an extra
component to the system that is not being modelled here (e.g., a
binary companion).  Long-baseline ($>$20m) data with closure-phase
arrays are needed to understand this better.

These diameter measurements allowed the stellar and dust contributions
to be separated in most cases.  By increasing the angular resolution
in the near-IR by 3-10$\times$ over best current literature
measurements, our new data offer strong constraints for new radiative
transfer modeling.  In addition, the dust shells for a few sources were
imaged using maximum entropy method.  When assisted by a MEM prior
incorporating the long-baseline IOTA data, dust shell asymmetries and
clumpiness are unambiguously identified and separated from
photospheric light.  Unfortunately, we were only able to confidently
make images for a subset of these sources, due to problems with
calibration of the atmospheric transfer function in the aperture
masking experiment and to incomplete sampling at longer baselines;
however, these limited results have proved enlightening.  Most
importantly, we have found a bipolar dust shell geometry for NML~Cyg,
as earlier suggested by OH, H$_2$O, and SiO masers, giving credence
to some alternative mass-loss mechanisms (e.g., involving
magnetic fields and/or rotation).

\begin{figure}[thb]
\begin{center}
%\epsscale{.5}
\includegraphics[angle=90,width=\columnwidth]{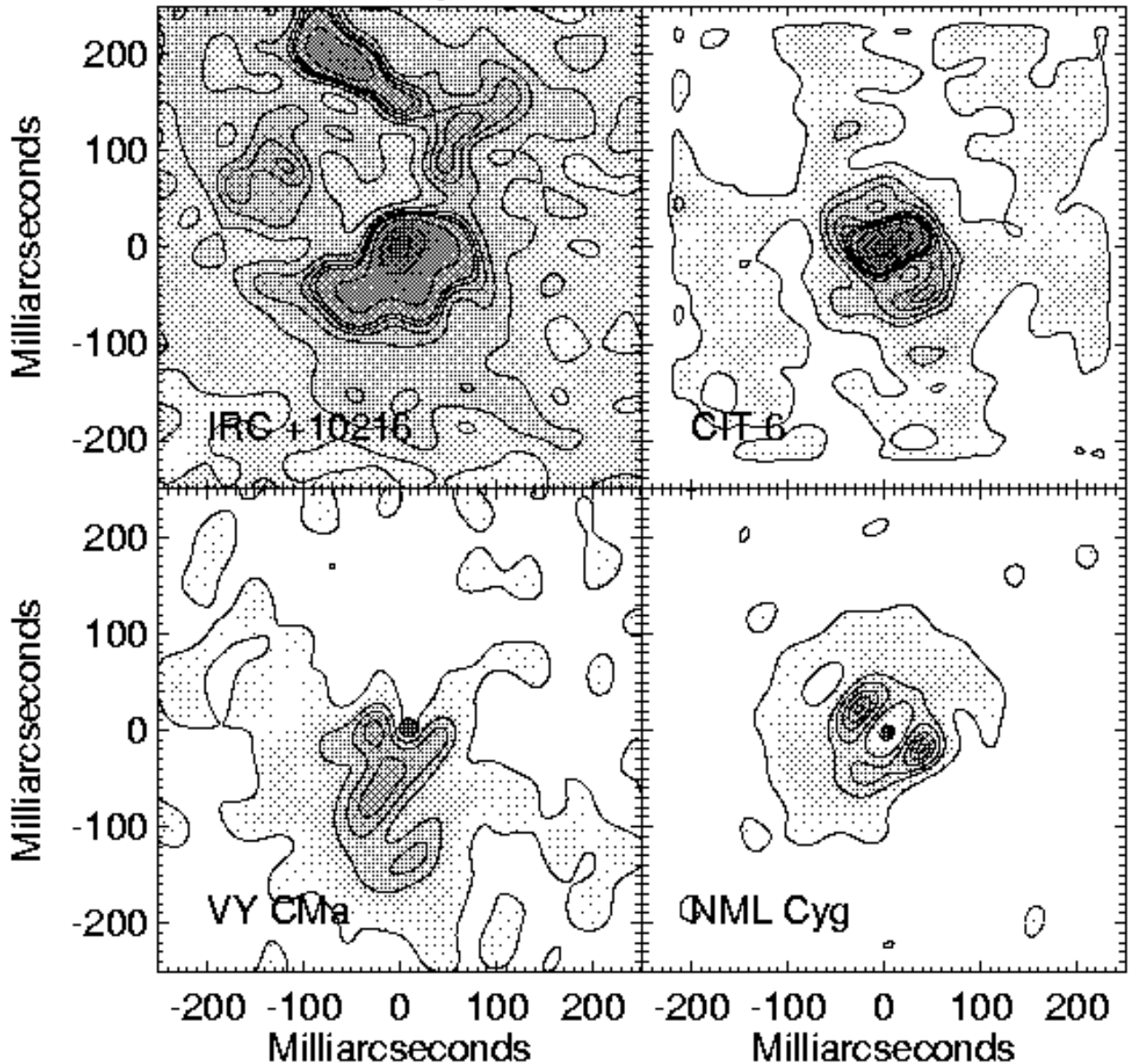}
\figcaption{This figure shows 2.2$\mu$m images of mass-losing
evolved stars reconstructed using Keck aperture masking data.
The images of the carbon stars 
IRC~+10216 and CIT~6 are from \citet{tuthill2000a} and
\citet{monnier_cit6} respectively, while the images
of VY~CMa and NML~Cyg are from this work and have incorporated
long-baseline IOTA data.  The contours levels are 0.1, 0.5, 1, 2, 3, 4, 5, 10, 30, 
and 70\% of the peak brightness in each panel, except for NML~Cyg where
the contours levels are 10$\times$smaller (owing to its fainter dust shell and more
compact central source).
Although circularly-symmetric dust shells likely exist, we have yet to image one 
successfully using optical interferometric techniques.
\label{gallery}}
\end{center}
\end{figure}

While it lies beyond of the scope of this paper, future detailed
modeling of the data presented here will dramatically improve our
knowledge of these sources and our results point the way toward new
classes of dust shell models.  It has lately been fashionable to
extend spherically-symmetric radiative transfer modeling up to (or
beyond) the range of applicability, by incorporating multiple dust
shells and unusual dust properties to fit multi-wavelength dust shell
observations.  When viewed together with other recent aperture masking
results (see Figure~\ref{gallery}), the new images presented here
strengthen the argument that clumpiness and global asymmetry should be
considered more seriously as the main explanation for the observed
deviations from simple uniform-outflow models.  We suggest that {\em
global} dust shell properties are best derived from mid-IR
observations where dust emission dominates over stellar and the
effects of clumpiness are better ``averaged-out'' by the intrinsically 
larger
emission volume in the mid-IR.  Presumably, the larger emission volume
will encompass many such ``clumps'' as well as a longer span of mass-loss
history, and should represent average dust shell properties more
faithfully.

We have also showed examples of how MEM imaging of interferometry data
can yield very different dust shell images, depending on the MEM prior
being used, and have discussed the difficulties in imaging faint dust
shells around bright stars.  We recognize and emphasize that optical
interferometry is still at an early stage of development, and recent
image reconstructions can not be interpreted as straightforwardly as
those derived from the Very Large Array (VLA) or other radio
interferometers.  The use of {\em a priori} information is critical
for accurately interpreting data from marginally resolved sources, and
new imaging software is needed to facilitate this (the method used
here was admittedly {\em ad hoc}).  All the visibility data here (Keck
masking and IOTA) have been converted to the new FITS format for
Optical Interferometry data (OI-FITS)
\footnote{http://www.mrao.cam.ac.uk/~jsy1001/exchange/} and are
available upon request.

\acknowledgments 
{Results from the IOTA interferometer would not have been possible
without continued support from the Smithsonian Astrophysical
Observatory.  This research has made use of the SIMBAD database,
operated at CDS, Strasbourg, France, and NASA's Astrophysics Data
System Abstract Service.  Some data presented herein were obtained at
the W.M. Keck Observatory, which is operated as a scientific
partnership among the California Institute of Technology, the
University of California and the National Aeronautics and Space
Administration.  The Keck Observatory was made possible by the
generous financial support of the W.M. Keck Foundation.  The {\em
getCal} program is a product of the Michelson Science Center (IPAC,
Caltech).}

\appendix
\section{IOTA-FLUOR Data Reduction Details}
\label{appendixa}
\subsection{Basic procedures}
Reduction of the FLUOR data was carried out using custom software
developed using the {\em Interactive Data Language} (IDL), similar in
its main principles to that described by \citet{foresto1997}.

The major steps of the data reductions are:
\begin{enumerate}

\item{Data inspection.  The raw fringe data is background-subtracted and
inspected.  Cosmic ray hits and other detector anomalies are automatically
detected and removed.  Visual inspection of the power spectra allow for
flagging of data corrupted by instrumental problems, in particular delay line
vibrations. Although troublesome, these problems are easily identified and removed
from the data stream.}
\item{``Kappa'' matrix.  \citet{foresto1997} described the use of the
kappa matrix for removal of photometric fluctuations and normalization
of the fringe amplitudes.  The kappa matrix is chromatic and thus must
be measured separately for each source and calibrator. Stability of
the kappa matrix during the night is a useful diagnostic of data
quality. }
\item{Removal of photometric fluctuations. During poor seeing, rapid
coupling fluctuations will contain high-frequency power which mimic
 real fringes.
The interferometric channels
have the incoherent part of the flux removed using the photometric signals and the
kappa matrix, which eliminates scintillation and coupling fluctuations (a strong effect).}
\item{Fringe normalization.  In each scan, the expected fringe
envelope for 100\% coherent light (unity visibility) is calculated
from the photometric and kappa measurements
\citep[e.g.,][]{foresto1997}, which allows for precise calibration of
the observed fringe visibility independent of the atmosphere.
\citet{foresto1997} advocate dividing the fringe data by the envelope
at this stage, however we have pursued a different strategy which is
more robust for low light levels and is described in the next section of the
appendix.}
\item{Power Spectra Measurement.  Next the power spectra are
calculated for each scan.  Noise sources cause a bias in the power
spectrum which must be removed.  The contribution from read noise is
estimated from calibration measurements of dark sky, while the
remaining bias (from photon noise and uncorrected scintillation) is
estimated by measuring the power at frequencies both above and below
the fringe frequency and interpolating for the intermediate
(fringe) frequencies; this bias term is subtracted for each scan.  For the
classical FLUOR analysis from \citet{foresto1997}, this ``power'' is
directly proportional the $V^2$ (Squared-Visibility) and can be
averaged.  In our method, we combine this measurement with the
normalization factor appropriate for that scan, and
make a weighted average of the normalized scans using bootstrap
sampling \citep{bootstrap}.}
\item{Instrumental Response. The above procedure is repeated for the
target and calibrator stars.  After correction for finite size
effects, the calibrator $V^2$ are used to monitor the instrumental
transfer function as a function of time.  Using simple linear
interpolation to estimate the transfer function at the times the
target observations were made, we divide the target $V^2$ by the
interpolated calibrator $V^2$ to yield a final calibrated measurement
of $V^2$.}
\item{Standard data quality checks.  We always analyze the two
interferometric channels independently and also apply both the
classical FLUOR method and our new normalization scheme in parallel.
Our results from the the two methods, and for both fringe outputs, are
statistically and internally consistent for bright sources. }
\end{enumerate}

\subsection{Normalization Scheme}
\label{normscheme}
Here, 
we describe more quantitatively the novel normalization 
procedure used in the IOTA-FLUOR data reduction.

The method of dividing by the fringe envelope, as described in detail
by \citet{foresto1997}, does correct for the varying photometric
signal strengths, but amplifies noise when the signals are small.  For
bright sources, signals are never small and thus this limitation poses
no problem; in fact, dividing by the envelope (sample-by-sample)
maximizes the precision of observations when limited by calibration of
coupling fluctuations (i.e., when you are not limited by detector
read-noise or photon-noise).  However in the experiment reported here,
we observed low visibility sources which had a signal-to-noise ratio
(in a single scan) which was sometimes below the threshold used by traditional
FLUOR analysis (SNR$\sim$3).

In our method, we measure an average normalization for each scan based
on the photometric signal.  Hence, rather than treating each scan
equally when taking the power spectrum, we have assigned a
normalization that is used to both weight the average power spectrum
and also allows the weighting to be done in a statistical way that is
free of bias.  Here we briefly describe the method \citep[see][for
more background on the notation and related methods]{foresto1997,
monnier2001}.

As already mentioned, the incoherent flux that appears on the
interferometric channels (I1, I2) is a linear combination of the
signals that appear on the photometric channels (P1, P2); the kappa
matrix can be used then to ``predict'' (I1,I2) given (P1,P2).  In
addition, we can use the components of the kappa matrix to predict the
maximum amplitude of the coherent part of the interferometric
channels.  This can written as:

\begin{equation}
I_1 = \kappa_{(P_1,I_1)} P_1 + \kappa_{(P_2,I_1)} P_2 + 
      2  \sqrt{\kappa_{(P_1,I_1)} P_1 \kappa_{(P_2,I_1)} P_2} \cdot \gamma(t)
\end{equation}

Here, $\gamma(t)$ is the mutual coherence function and encodes the
fringe visibility, the quantity we wish to measure.  Usually $\gamma$ is
temporally modulated by adjusting the relative path lengths in the two
arms of the interferometer.  An equation for $I_2$ follows from the above.
Hence, for perfect coherence $\|\gamma\|$=1, the maximum measured fringe amplitude
would be ({\em not} normalized by mean flux):
\begin{equation}
I_1^{\rm envelope} = 2 \sqrt{\kappa_{(P_1,I_1)} P_1 \kappa_{(P_2,I_1)}
P_2}
\end{equation}

In the power spectrum method, the $V^2$ is measured because it is free
of bias from read noise and photon noise.  Applying Parseval's Theorem
to the coherent part of the fringe interferogram, 
we can understand this integration of
the fringe ``power'' in the Fourier (frequency) Space as equivalent to
the integration of the square of the fringe envelope in time.
Hence, we intend to normalize the measured fringe ``power'' by the
average of:

\begin{equation}
 (I_1^{\rm envelope})^2 = 4 \kappa_{(P_1,I_1)} \kappa_{(P_2,I_1)} P_1 P_2
\end{equation}

\begin{figure}[thb]
\begin{center}
%\epsscale{.5}
\includegraphics[angle=90,height=3in]{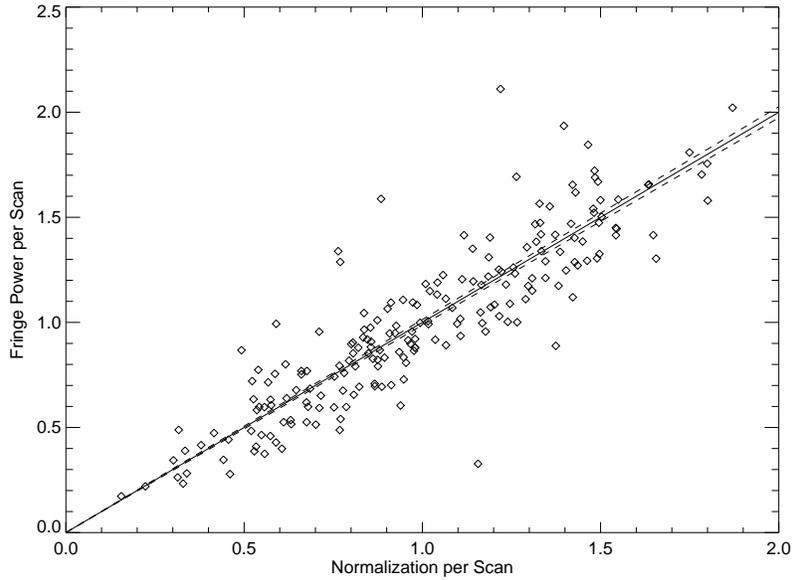}
\figcaption{Normalization Scheme.  This figure illustrates the new
calibration method being employed for analyzing IOTA-FLUOR data.  The
(bias-corrected) fringe power observed in each fringe scan (of 200) is
plotted against a ``normalization'' factor estimated from the
photometric channels of FLUOR.  The $V^2$ is simply proportional to
the slope of this relation, which is plotted here along with its
uncertainty.  For this single dataset, the formal uncertainty in the
slope is 1.3\% in $V^2$, which is only 0.65\% for the Visibility.
\label{normeg}}
\end{center}
\end{figure}

Figure~\ref{normeg} shows an example of this method applied to one
interferometric output of a single observation set of a bright source
($\sigma$~CMa).  This figure shows that the (bias-corrected) fringe
power varies by a large amount due to coupling fluctuations but that
our ``normalization'' factor, calculated from the photometric
channels, faithfully tracks this variation.  We can also see the
fringe power is very linear with normalization and a simple
weighted-mean is used to estimate the slope of the relation, a value
proportional to $V^2$.  In this example, the error was calculated
using a bootstrap method and the 1.3\% uncertainty in the slope is
reflected in the plot.

Typically, one might worry that Poisson and read noise would bias the
above normalization.  However, one can see that this quantity is not
biased as long as $P_1$ and $P_2$ are uncorrelated on the timescale of
single scan ($<<1$ sec), a reasonable assumption for a long-baseline
interferometer where the atmospheric distortion above an aperture is
independent of the others.  Note, that this useful statistical property also
implies that one could also average $P_1$ and $P_2$ separately,
as pointed out by \citet{shaklan1992}, and still have a good bias-free
estimate of the fringe amplitude normalization.

\bibliographystyle{apj}
\bibliography{apj-jour,KeckIOTA,Thesis,Review,Review2}

\clearpage
\begin{deluxetable}{lllllll}
\tablecaption{Basic Properties of Targets\label{targets}}
\scriptsize
%\tabletypesize{\scriptsize}
\tablewidth{0pt}
\tablehead{
\colhead{Target} & \colhead{RA (J2000)} & \colhead{Dec (J2000)} &
\colhead{V} & \colhead{K} &
\colhead{Spectral} & \colhead{Type of Source}\\
\colhead{Name} & & &\colhead{mag\tablenotemark{a}} & \colhead{mag\tablenotemark{a}} & \colhead{Type} & 
}
\startdata
IK~Tau  &  03 53 28.84 & $+$11 24 22.6 & 11.9 & -1.1 & M10III & Dust-enshrouded Mira variable \\
VY~CMa & 07 22 58.33 &  $-$25 46 03.2 & 8.0 & +0.1 & M3/4I & Dust-enshrouded red supergiant \\
HD~62623 & 07 43 48.47 & $-$28 57 17.4 & 4.0 & 2.3 & A3Iab & A supergiant with
infrared excess \\
R~Leo & 09 47 33.49 &  $+$11 25 43.6 & 6.0 & -2.2 & M8III & Mira variable \\
RT~Vir & 13 02 37.98 & $+$05 11 08.4 & 8.6 & -1.0 & M8III & Semi-regular variable \\
R~Hya & 13 29 42.78 & $-$23 16 52.8  & 6.4 & -2.6 & M7III & Mira variable \\
W~Hya & 13 49 01.00 & $-$28 22 03.5  & 7.5 & -3.1 & M7III & Mira variable \\

VX~Sgr & 18 08 04.05 & $-$22 13 26.6 & 10.0 & 0.2 & M4-9.5I & Dust-enshrouded red superigant \\
IRC~+10420 & 19 26 48.03 & $+$11 21 16.7 & 8.5 & 3.6 & F8I & Rapidly-evolving F supergiant \\
NML~Cyg & 20 46 25.46 & $+$40 06 59.6 & 16.6 & 0.6 & M6I & Dust-enshrouded red supergiant \\

\enddata
\tablenotetext{a}{Most of the targets are variable stars and these
magnitudes are merely representative.  See text for more recent
photometry.}
\end{deluxetable}

\clearpage

\begin{deluxetable}{llllll}
%\tabletypesize{\scriptsize}
\scriptsize
\tablecaption{Journal of Observations \label{observations}}
\tablewidth{0pt}
\tablehead{
\colhead{Target} & \colhead{Date}   & \colhead{$\lambda$}   &
\colhead{$\Delta \lambda$} & \colhead{Interferometer}  & \colhead{Calibrator}\\
\colhead{(Sp. Type)} & \colhead{(UT)}   & \colhead{($\mu$m)}   &
\colhead{($\mu$m)} & \colhead{Configuration}  & \colhead{Names}
}
\startdata
HD~62623        & 1999Apr26 & 2.269& 0.155& Keck Annulus & HD~47667 \\
(A3Iab)         & 2000Jan26 & 2.269& 0.155& Keck Annulus & 54~Per \\
                & 2000Feb05,07 & 2.16 & 0.32 & IOTA N15-S15 & HD~63852 \\
                & 2000Feb10 & 2.16 & 0.32 & IOTA N25-S15 & HD~63852 \\
\tableline
IK~Tau   & 2000Jan26 & 2.257& 0.053& Keck Golay 21 & $\alpha$~Cet\\
(M10III)                & 2000Feb03,05,07 & 2.16 & 0.32 & IOTA N15-S15 & $o$~Tau \\
                & 2000Feb08,10 & 2.16 & 0.32 & IOTA N25-S15 & $o$~Tau \\
\tableline       
IRC~+10420 & 2000Jun16,17 & 2.16 & 0.32 & IOTA N35-S15 & SAO~104467 \\
(F8I)                   & 2000Jun24 & 2.269& 0.155& Keck Annulus & SAO~104655,SAO~104467\\
\tableline
NML~Cyg         & 2000Jun23 & 2.16 & 0.32 & IOTA N35-S15 & SAO~49410 \\
(M6I)           & 2000Jun24 & 2.257& 0.053& Keck Annulus & $\xi$~Cyg\\
                & 2000Jun24 & 2.257& 0.053& Keck Golay 21& SAO~105500\\
                & 2000Jun24 & 2.269& 0.155& Keck Golay 21& $\xi$~Cyg \\
\tableline
R~Hya           & 2000Jan26 & 2.257& 0.053& Keck Golay 21& $\delta$~Vir, 2~Cen, $\pi$~Leo, SW~Vir\\
(M7III)         &2000Feb04,05 & 2.16 & 0.32 & IOTA N15-S15 & $\gamma$~Hya \\
               & 2000Feb08,09 & 2.16 & 0.32 & IOTA N25-S15 & $\gamma$~Hya \\
\tableline
R~Leo   &       2000Jan26 & 2.257& 0.053& Keck Golay 21& $\pi$~Leo \\
(M8III)         &2000Feb01,05,06 & 2.16 & 0.32 & IOTA N15-S15 & $\pi$~Leo \\
\tableline
RT~Vir          &2000Feb03,05 &  2.16 & 0.32 & IOTA N15-S15 & $\sigma$~Vir \\
(M8III)         &2000Apr12,13,15&  2.16 & 0.32 & IOTA N35-S15 &  $\sigma$~Vir  \\
                &2000Apr20 & 2.16 & 0.32 & IOTA N25-S15 &   $\sigma$~Vir \\
\tableline
VX~Sgr & 2000Apr12,13,14,20 & 2.16 & 0.32 & IOTA N35-S15 & SAO~186841 \\
(M4-9.5I)               & 2000Apr23 & 2.16 & 0.32 & IOTA N25-S15 & SAO~186841 \\ 
                & 2000Jun21 & 2.16 & 0.32 & IOTA N35-S15 & SAO~186841 \\
                & 2000Jun24 & 2.249& 0.024& Keck Annulus & 14~Sgr \\
                & 2000Jun24 & 2.257& 0.053& Keck Golay 21& SAO~186681, SAO~186841\\
\tableline
W~Hya           & 2000Jan26 & 2.249& 0.024&Keck Golay 21 & 2~Cen \\
(M7III) & 2000Feb04,05,06 &  2.16 & 0.32 & IOTA N15-S15 & $\pi$~Hya \\
\tableline
VY~CMa  & 1999Feb05 & 2.257& 0.053& Keck Golay 21& $\sigma$~CMa \\
(M3/4I)         & 2000Jan26 & 2.269& 0.155& Keck Annulus & 54~Per, $\pi$~Leo, $\sigma$~CMa \\
                & 2000Feb04,05,07 & 2.16 & 0.32 & IOTA N15-S15 & $\sigma$~CMa \\
                & 2000Feb08,10 & 2.16 & 0.32 & IOTA N25-S15 & $\sigma$~CMa              
 \enddata

%% Text for table notes should follow after the \enddata but before
%% the \end{deluxetable}. Make sure there is at least one \tablenotemark
%% in the table for each \tablenotetext.

%\tablenotetext{a}{Sample footnote for table~\ref{tbl-1} that was generated
%with the deluxetable environment}
%\tablenotetext{b}{Another sample footnote for table~\ref{tbl-1}}%
%
%\tablecomments{Occasionally, authors wish to append a short
%paragraph of explanatory notes that pertain to the entire table, but
%which are different than the caption.  Such notes should be placed in
%a {\tt tablecomments} command like this.}
%
\end{deluxetable}

\clearpage
\begin{deluxetable}{llll}

%\tabletypesize{\scriptsize}
\scriptsize
\tablecaption{Calibrator Information\label{calibrators}
($\ast$ indicates a calibrator of long-baseline IOTA interferometer data
where accurate diameters are most critical).
}
\tablewidth{0pt}
\tablehead{
\colhead{Calibrator} & \colhead{Spectral} & \colhead{Adopted Uniform Disk\tablenotemark{a}}  &
\colhead{Reference(s)}\\ 
\colhead{Name} & \colhead{Type} & \colhead{Diameter (mas)} &   }

\startdata
2~Cen & M4.5III & 13.9 $\pm$1.4 & \citet{heras2002} \\
14~Sgr & K2III & 2.3$\pm$1.8  & getCal\tablenotemark{b} \\
54~Per & G8III & 1.41$\pm$0.13  & CHARM\tablenotemark{c}  \\
$\alpha$~Cet & M1.5III & 11.6$\pm$0.4 & CHARM \\
$\delta$~Vir & M3III & 10.7 $\pm$ 1.0 & \citet{heras2002}\\
$\gamma$~Hya$^\ast$ & G8III & 3.4 $\pm$0.5 & getCal \\
HD~47667        & K2III & 2.56$\pm$0.04 & CHARM \\
HD~63852$^\ast$ & K5III & 2.3$\pm$1.7 & getCal \\
$o$~Tau$^\ast$ & G6III (SB) & 2.7$\pm$0.3 & CHARM, CADARS\tablenotemark{d} \\
$\pi$~Hya$^\ast$ & K2III & 3.7$\pm$0.1 & CHARM \\
$\pi$~Leo$^\ast$ & M2III  & 4.85$\pm$0.23 & CHARM \\
SAO~49410$^\ast$ & K5Iab &2.9$\pm$0.5 & \citet{vanbelle1999}\tablenotemark{e} \\
SAO~104467$^\ast$ & K0V & 1.7$\pm$0.3 & getCal \\ SAO~104655 &
G8II-III & 1.5$\pm$0.2 & getCal\\ SAO~105500 & M0III & 5.5$\pm$0.5 &
CHARM \\ 
SAO~186681 & K3III & 6.9$\pm$0.9 & getCal\\
SAO~186841$^\ast$ & K1III & 4.4$\pm$0.2 & CHARM\\ 
SW~Vir & M7III & 16.81$\pm$0.12 & CHARM \\ 
$\sigma$~CMa$^\ast$ & M0Iab & 8.9$\pm$1.2 & getCal\\ 
$\sigma$~Vir$^\ast$& M2III & 6.2$\pm$1.0 & getCal \\
$\xi$~Cyg & K4.5I & 6.0$\pm$1.3 & getCal, CHARM\\ \enddata
\tablenotetext{a}{The diameter error quotes have not been validated
independently.  While adequate for our purposes here, workers who
require precision calibration are warned to research their calibrators
carefully and not rely too heavily on ``catalogs'' such as CHARM.}
\tablenotetext{b}{{\em getCal} is maintained and distributed by the
Michelson Science Center (http://msc.caltech.edu)}
\tablenotetext{c}{CHARM is the Catalog of High Angular Resolution
Measurements \citep{richichi2002}} 
\tablenotetext{d}{CADARS is the
Catalog of Apparent Diameters and Absolute Radii of Stars
\citep{fracassini2001}} 
\tablenotetext{e}{The diameter recorded in
this reference is in error; however, the reported $V^2$ measurement
is correct (van Belle, 2003, private communication) 
and we have used this to calculate the UD diameter found herein.}
\end{deluxetable}

%% If you use the table environment, please indicate horizontal rules using
%% Do not put multiple tabular environments within a single table.
%% The optional \label should appear inside the \caption command.

\clearpage
\begin{deluxetable}{lll}

%\tabletypesize{\scriptsize}
\tablecaption{Results of Uniform Disk Fits\label{diameters}
}
\tablewidth{0pt}
\tablehead{
\colhead{Source} & \colhead{Uniform Disk} &\colhead{Comments} \\
\colhead{Name} & \colhead{Diameter\tablenotemark{a} (mas)}  &   }
\startdata
IK~Tau& 20.2$\pm$0.2$\pm$0.3 & Large deviation from Uniform Disk \\
& & See text for other provisos \\
NML~Cyg& 7.8$\pm$0.4$\pm$0.5 & Unexpectedly small; large gap in baseline coverage \\
R~Hya & 23.7$\pm$0.8$\pm$0.6 & Some deviation from Uniform Disk ($\chi^2\sim2$)\\ 
R~Leo & 30.3$\pm$0.2$\pm$0.3 & Excellent fit to Uniform Disk\\
RT~Vir & 12.4$\pm$0.1$\pm$0.4 & Good fit  \\
VX~Sgr & 8.7$\pm$0.3$\pm$0.1 & Excellent fit (IOTA data only) \\
VY~CMa & 18.7$\pm$0.3$\pm$0.4 & Good fit; star contributes 36\% of Kband flux  \\
W~Hya & 42.5$\pm$0.7$\pm$0.4 & Poor fit ($\chi^2\sim5$) \\
\enddata
\tablenotetext{a}{The best-fit UD diameter (for $\lambda_{\rm eff}=2.16\mu$m)
is followed by estimates
of the statistical error, then by an estimate of the systematic error
(see text in  \S\ref{validation}). The
systematic error is usually dominated by uncertainty in the
calibrator diameter.
We emphasize that we measure
an apparent diameter, and relation to the true
``photospheric'' diameter relies on additional assumptions of
limb-darkening and other effects; some late-type stars are known
to have different apparent sizes between the visible, near-IR and
mid-IR wavelengths \citep[e.g.,][]{weiner2000}.
}
\end{deluxetable}

\clearpage

\clearpage

\clearpage

\clearpage

\clearpage

\clearpage

\clearpage

\clearpage

\clearpage

\clearpage

\clearpage

\clearpage

\clearpage

\clearpage

\clearpage

\clearpage

\clearpage

\clearpage

\clearpage

\clearpage

\clearpage

\clearpage

\clearpage

\end{document}